\documentclass[aps,pre,preprint,superscriptaddress]{revtex4-2}

\usepackage{amsmath,amssymb,amsthm}
\usepackage{graphicx}
\usepackage{bm}
\usepackage{hyperref}
\usepackage{color}

\begin{document}

\title{
Temporal Coarse-Graining of Latent Default-Probability Paths Generates Effective Default Correlation}

\author{Shintaro Mori}
\affiliation{Graduate School of Science and Technology, Hirosaki University}

\begin{abstract}
We show that persistent dynamics of a latent default-probability path can generate effective default correlation through temporal coarse-graining. In the OU--Binomial baseline, monthly defaults are conditionally independent given this latent path, but aggregating monthly default probabilities into long-horizon probabilities induces a scale-dependent effective mixing distribution for aggregated default counts. Applied to corporate default-count data, this mechanism explains long-horizon overdispersion, autocorrelation, and the emergence of effective default correlation. We then examine Davis--Lo-type contagion and Vasicek-type common-factor extensions. Direct fitting at each aggregation scale assigns increasing residual covariance shares to instantaneous dependence, but worsens the per-block expected log predictive density. In contrast, when monthly posterior latent paths are first coarse-grained and residual-dependence parameters are estimated conditional on these paths, the residual covariance contributions remain small while the predictive density improves. Thus, temporal coarse-graining provides a scale-consistent baseline that regularizes the attribution of variance and improves identifiability by suppressing the over-allocation of long-horizon fluctuations to contagion or asset-correlation parameters.
\end{abstract}


\maketitle

\section{Introduction}

In econophysics and sociophysics, macroscopic phenomena are often interpreted as
emergent outcomes of interactions among heterogeneous agents
\cite{MantegnaStanley1999,Galam2008,Lux1995,LuxMarchesi1999,
Alfarano2005,Bouchaud2002,FernandezGracia2014,
MoriHisakadoTakahashi2012,Mori2019}. Yet similar aggregate patterns can also be
generated by time-varying aggregate states, without specifying the underlying
microscopic interactions \cite{SmolyakHavlin2022}. 
This distinction is particularly important in credit-risk data, where default 
clustering may reflect direct interactions among firms, common-factor dependence, 
or persistent fluctuations in default risk
\cite{Schonbucher2003,DavisLo2001,Vasicek1991,Vasicek2002,
DasDuffieKapadiaSaita2007,DuffieEcknerHorelSaita2009,
AzizpourGieseckeSchwenkler2018}. 
When default counts are observed only after temporal aggregation,
understanding how dynamical fluctuations in default risk are transformed into
apparent cross-sectional dependence becomes essential for both model
identification and risk management.

Default clustering has been modeled through several complementary mechanisms.
Contagion models attribute clustering to direct or indirect interactions among
firms \cite{DavisLo2001,SakataHisakadoMori2007,TorriGiacomettiFarina2026,
HisakadoMori2021}, while common-factor and frailty models attribute it to
correlated exposure to aggregate risk factors
\cite{Vasicek1991,Vasicek2002,HisakadoMori2021,Schonbucher2003,
DasDuffieKapadiaSaita2007,DuffieEcknerHorelSaita2009,
AzizpourGieseckeSchwenkler2018}. Event-time and self-exciting point-process
approaches provide another important framework for default clustering when
detailed timing information is available
\cite{Hawkes1971,ErraisGieseckeGoldberg2010,Kirchner2017,
HisakadoHattoriMori2022}. In aggregated default-count data, however, these
mechanisms are difficult to distinguish because only the total number of
defaults over a given observation period is observed.
In previous work, this identifiability problem was examined by comparing the
loss distributions generated by contagion and common-factor models, and by
measuring their separation using information-theoretic distances such as the
Kullback--Leibler divergence \cite{Mori2026ContagionMacro}.
Such distributional comparisons clarify whether different mechanisms leave
distinct signatures in aggregated counts. They also reveal a limitation: when
the observation period is long and the number of annual observations is small,
model identification based on empirical annual distributions can be sensitive to
unresolved persistent fluctuations in the latent default probability.
The present study clarifies the origin of this distributional identifiability
problem from the viewpoint of temporal coarse-graining.

In this paper, we address this limitation by treating the observation horizon
itself as a coarse-graining scale. We first infer a monthly latent
default-probability path and then coarse-grain this path to longer horizons
through survival-based temporal aggregation. This construction does not fit an
independent mixture distribution at each horizon. Instead, it transforms the
temporally correlated monthly latent path into a scale-dependent effective
mixing distribution of the coarse-grained default probability. This effective
mixing distribution induces the long-horizon default-count distribution and the
corresponding effective default correlation. Thus, same-period default
correlation inferred at a longer horizon is treated as an effective quantity
shaped by temporal coarse-graining, rather than as a direct measure of
microscopic instantaneous dependence.
 
We demonstrate this interpretation by comparing two modeling routes. In the
direct-fitting route, each aggregation scale is fitted independently. This
allows unresolved within-horizon dynamics of the latent default probability to
be absorbed by Davis--Lo-type contagion or Vasicek-type common-factor
parameters, producing an apparent residual covariance component. However, this
additional flexibility worsens the per-block elpd relative to the OU--Binomial
baseline, suggesting that direct fitting can over-allocate variance to
same-period dependence, especially when the number of long-horizon observations
is small.

In the renormalized route, residual dependence is first estimated at the
monthly scale, and the resulting posterior latent default-probability paths are
then coarse-grained to longer horizons. Residual-dependence parameters are
estimated only after conditioning on these coarse-grained latent paths. This
construction suppresses the over-allocation of long-horizon variance to residual
covariance parameters. The residual covariance share remains small, whereas the
renormalized OU--Davis--Lo and OU--Vasicek specifications improve the per-block
elpd relative to the corresponding OU--Binomial baselines. 
The improved predictive density, together with the small residual covariance
shares, indicates that temporal coarse-graining provides a scale-consistent
baseline that accounts for most of the variance through the coarse-grained
latent path, while a small residual instantaneous-dependence component only
refines the shape of the predictive count distribution.

The identifiability problem arises because contagion, common-factor dependence,
and latent default-probability fluctuations can generate similar aggregated
count distributions. Under direct fitting, this ambiguity allows unresolved
long-horizon fluctuations to be reabsorbed into instantaneous-dependence
parameters, leading to an over-allocation of variance to residual covariance
components. By first constructing a scale-consistent
coarse-grained latent path and then estimating residual-dependence parameters
conditional on that path, the renormalized route separates these two sources of
variation more conservatively. 
In this sense, temporal coarse-graining provides both a diagnostic explanation
of the identifiability problem and a practical way to improve identifiability
across observation scales.

The remainder of this paper is organized as follows.
Section II introduces a binomial state-space model with an Ornstein--Uhlenbeck
(OU) latent default-probability process and defines the temporal
coarse-graining procedure for posterior latent paths. It also presents the
empirical variance, autocorrelation, and effective-mixing diagnostics for the
OU--Binomial baseline.
Section III introduces the OU--Davis--Lo and OU--Vasicek residual-dependence
extensions, describes the direct and renormalized fitting routes, and compares
their covariance-share and predictive-density diagnostics. 
Section IV summarizes the implications for interpreting default correlation in
aggregated default-count data.

\section{OU-Binomial model and temporal coarse-graining}

This section introduces the baseline dynamical model used to represent
persistent dynamics in monthly default-count data.
We formulate a monthly OU--Binomial state-space model, in which defaults are
conditionally independent and binomially distributed given a latent monthly
default probability. The latent
default-probability process is modeled as a persistent Gaussian process on the
probit scale. This model contains no instantaneous default correlation or
contagion component; all overdispersion beyond conditional binomial noise is
therefore attributed to temporal fluctuations of the latent
default-probability path.

We then define temporal coarse-graining as a mapping from monthly posterior
latent paths to coarse-grained default probabilities over longer observation
horizons. This construction transforms temporal persistence in the monthly
latent path into a scale-dependent effective mixing distribution of
coarse-grained default probabilities, without re-estimating a separate mixture
model at each aggregation scale. The induced mixing distribution then generates
the long-horizon default-count distribution and provides the baseline for
interpreting same-period default correlation as a coarse-grained effective
quantity.

The purpose of this section is threefold. First, we show how monthly latent
default-probability paths generate long-horizon default-count distributions
through survival-probability aggregation. Second, we examine whether the
coarse-grained latent path explains the aggregation-scale dependence of the
monthly-equivalent default-rate variance and autocorrelation. Third, we analyze
the induced effective mixing distribution and compare it with a time-shuffled
benchmark in order to isolate the role of temporal persistence in the latent
default-probability path.

\subsection{Monthly OU--Binomial state-space model}

Let $L_t$ denote the number of defaults observed in month $t$, and let $n_t$
denote the corresponding number of obligors. As the baseline OU--Binomial
model, 
we assume that individual defaults occur independently conditional on a latent
monthly default probability $p_t$.
The observed default count then follows
\[
L_t \mid p_t \sim \mathrm{Binomial}(n_t,p_t).
\]
Here, $p_t$ is interpreted as the latent monthly average default probability.
It summarizes aggregate variation in default conditions at month $t$ without
identifying a separate macroeconomic factor.

The latent probability $p_t$ is modeled on the probit scale. Specifically, we
write
\[
y_t = \Phi^{-1}(p_t),
\qquad
p_t = \Phi(y_t),
\]
where $\Phi$ denotes the cumulative distribution function of the standard normal
distribution. To describe persistent fluctuations of the latent default
probability, we assume that $y_t$ follows a stationary first-order
autoregressive process,
\[
y_t - \mu = \phi (y_{t-1}-\mu) + \epsilon_t,
\qquad
\epsilon_t \sim N(0,\sigma_\epsilon^2).
\]
Equivalently, this process can be viewed as a discrete-time Ornstein--Uhlenbeck
process. The stationary variance of $y_t$ is
\[
\sigma^2 = \frac{\sigma_\epsilon^2}{1-\phi^2}.
\]
Thus, $\mu$ controls the long-run level of the latent default probability,
$\sigma$ measures the stationary magnitude of its fluctuations, and $\phi$
controls their temporal persistence.

This specification is intentionally minimal. It contains no instantaneous
default correlation, contagion, or asset-correlation term. Therefore, any
overdispersion explained by this model is attributed to time variation in the
latent default-probability path rather than to conditional dependence among
defaults within the same month. This makes the OU--Binomial model a natural
baseline for separating persistent latent-probability dynamics from residual
instantaneous dependence.

The monthly OU--Binomial model is estimated in a Bayesian framework. The
posterior distribution provides sample paths $\{p_t^{(s)}\}$ of the latent
monthly default probability. These posterior sample paths are then
coarse-grained to longer horizons through the temporal coarse-graining
procedure described below.

\subsection{Temporal coarse-graining of posterior latent paths}

Given posterior sample paths $\{p_t^{(s)}\}$ obtained from the monthly
OU--Binomial model, we next construct default-count distributions over longer
observation horizons. Let $k$ denote the aggregation scale measured in months.
For each non-overlapping block $b$, consisting of months
$t=bk,\ldots,bk+k-1$, we define the $k$-month default probability by aggregating
monthly survival probabilities:
\[
p_b^{(k)}
=
1-\prod_{j=0}^{k-1}\{1-p_{bk+j}\}.
\]
This relation follows from the probability that an obligor survives all $k$
months in the block. If the monthly default probabilities within the block are
$p_{bk},\ldots,p_{bk+k-1}$, then the probability of no default during the block
is $\prod_{j=0}^{k-1}(1-p_{bk+j})$, and the corresponding probability of at
least one default is $p_b^{(k)}$.

For each block, the corresponding number of obligors and defaults are denoted by
$n_b^{(k)}$ and $L_b^{(k)}$, respectively. In the empirical implementation,
$L_b^{(k)}$ is obtained by summing monthly default counts within the block,
whereas $n_b^{(k)}$ is taken as the number of obligors at the beginning of the
block, $n_b^{(k)}=n_{bk}$. Thus, the block-level default rate is interpreted as
an exposure-normalized block default count based on the initial block exposure.
Details of this construction and its consistency with the reported annual
S\&P series are given in Sec.~\ref{app:data_aggregation}.
Conditional on the coarse-grained default probability $p_b^{(k)}$, the
baseline long-horizon count model is
\[
L_b^{(k)} \mid p_b^{(k)}
\sim
\mathrm{Binomial}(n_b^{(k)},p_b^{(k)}).
\]

The important point is that $p_b^{(k)}$ is not estimated separately at each
aggregation scale. Instead, it is induced by the monthly posterior latent path.
For posterior sample $s$, we compute
\[
p_b^{(k,s)}=1-\prod_{j=0}^{k-1}\{1-p_{bk+j}^{(s)}\},
\]
where $p_t^{(s)}$ denotes the posterior sample path of the monthly default
probability. The collection $\{p_b^{(k,s)}\}_{b,s}$ then defines an empirical
effective mixing distribution at scale $k$, which we denote by $G_k$.
Because the monthly path is temporally persistent, $G_k$ is not determined only
by the one-month marginal distribution of $p_t$, but also by the temporal
ordering of high- and low-default-probability states.

Thus, temporal aggregation induces a scale-dependent effective mixing
distribution of the coarse-grained default probability, and hence the
long-horizon binomial mixture distribution
\[
P(L_b^{(k)}=\ell)=\int\binom{n_b^{(k)}}{\ell}p^\ell
(1-p)^{n_b^{(k)}-\ell}\,dG_k(p).
\]
This mixture distribution is different from a static mixture fitted directly to
$k$-month or annual data. It is generated by coarse-graining the monthly
posterior latent path, and therefore retains information about its temporal
persistence. In this sense, the mapping from $\{p_t\}$ to
$\{p_b^{(k)}\}$ is the temporal coarse-graining transformation studied in this
paper: temporal persistence in the monthly latent default-probability dynamics
is converted into an effective mixture at the longer observation horizon.

This construction provides a baseline against which residual instantaneous
dependence, such as contagion or common-factor default correlation, can be
evaluated. In the following subsections, we examine whether the temporally
coarse-grained latent default-probability path alone can account for the
aggregation-scale dependence of the monthly-equivalent default-rate variance,
the autocorrelation structure, and the induced long-horizon mixing distribution.

As a diagnostic, the same construction also gives a simple variance decomposition
for the monthly-equivalent default rate used in the empirical analysis. For the
OU--Binomial model, conditional on the coarse-grained default probability
$p_b^{(k)}$ and exposure $n_b^{(k)}$, the monthly-equivalent default rate
\[
r_b^{(k)}=\frac{L_b^{(k)}}{k n_b^{(k)}}
\]
has conditional mean and variance
\[
\mathbb{E}\!\left(r_b^{(k)}\mid p_b^{(k)},n_b^{(k)}\right)
=\frac{p_b^{(k)}}{k},
\qquad
\mathrm{Var}\!\left(r_b^{(k)}\mid p_b^{(k)},n_b^{(k)}\right)
=\frac{p_b^{(k)}\{1-p_b^{(k)}\}}{k^2 n_b^{(k)}}.
\]
For each posterior sample path $s$, the posterior predictive variance at scale
$k$ is evaluated over non-overlapping blocks as
\[
\mathrm{Var}_{\mathrm{post.pred.}}^{(s)}(r^{(k)})=
\frac{1}{B_k}\sum_{b=1}^{B_k}
\frac{p_b^{(k,s)}\{1-p_b^{(k,s)}\}}{k^2 n_b^{(k)}}+\mathrm{Var}_{b}\!\left(\frac{p_b^{(k,s)}}{k}\right),
\]
where $B_k$ is the number of $k$-month blocks and $\mathrm{Var}_{b}$ denotes the
sample variance over blocks. The reported components are posterior medians over
$s$.

\subsection{Data and empirical variance scaling}

We use monthly corporate default-count data from Standard \& Poor's (S\&P)
covering the period from January 1981 to September 2021 as the primary dataset
for estimating the latent default-probability dynamics.
For each month $t$, the data contain the number of defaults $L_t$ and the
corresponding number of obligors $n_t$. From these monthly observations, we
construct non-overlapping $k$-month aggregated counts for
$k=1,2,3,4,6,12$, using the block construction defined above.
Details of the exposure convention and its consistency with the reported annual
S\&P series are given in Sec.~\ref{app:data_aggregation}.

As external annual benchmarks, we use Moody's and S\&P annual default-count data
covering the period 1981--2023, following the annual analysis in the previous
study~\cite{Mori2026ContagionMacro}. These annual series provide reference
points for comparing the annual fluctuations implied by the temporally
aggregated monthly S\&P series; see Sec.~\ref{app:data_aggregation} for the
consistency check of the $k=12$ aggregation.

\begin{figure}[htbp]
    \centering
    \includegraphics[width=0.75\linewidth]{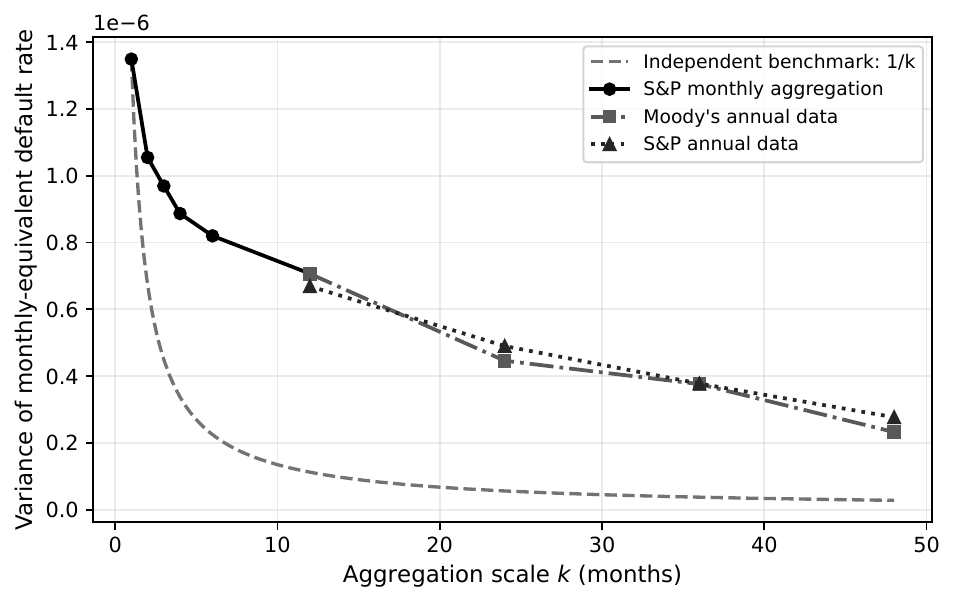}
    \caption{
Empirical variance scaling of the monthly-equivalent default rate.
The dashed line shows the independent benchmark proportional to $1/k$,
normalized at $k=1$. Annual Moody's and S\&P data for 1981--2023 are shown as
external benchmarks.
}
\label{fig:data_variance_scaling}
\end{figure}

For each aggregation scale $k$, we define the monthly-equivalent default rate by
\[
r_b^{(k)}=\frac{1}{k}\frac{L_b^{(k)}}{n_b^{(k)}}.
\]
This normalization puts all aggregation scales on a common monthly scale.
Figure~\ref{fig:data_variance_scaling} shows the empirical variance of
$r_b^{(k)}$ as a function of $k$.

If monthly default probabilities were independent over time, temporal
aggregation would average out fluctuations and the variance of the
monthly-equivalent default rate would decrease approximately in proportion to
$1/k$. The empirical scaling is substantially slower than this independent
benchmark. This indicates that fluctuations in the latent default probability
are persistent over time and cannot be treated as independent monthly noise.

At the annual scale, the variance obtained by aggregating the monthly S\&P data
to $k=12$ months is close to the variances observed in the annual Moody's and
S\&P benchmark series. The corresponding mean annual default rates are also of
the same order: approximately $1.50\%$ for the monthly S\&P series aggregated to
one year, $1.59\%$ for Moody's annual data, and $1.48\%$ for S\&P annual data.
The annual volatilities are likewise close, around one percentage point. These
comparisons support the use of the monthly S\&P series as a basis for
constructing temporally coarse-grained annual default-count distributions.

The slow decay of the variance is the empirical motivation for the
OU--Binomial state-space model introduced above. A persistent latent
default-probability process can generate long-horizon overdispersion through
temporal aggregation alone, before introducing a separate same-period
default-correlation parameter. In the next subsection, we estimate the monthly
OU--Binomial model and examine whether its posterior latent paths reproduce the
aggregation-scale dependence of the monthly-equivalent default-rate variance
and autocorrelation structure.

\subsection{OU--Binomial baseline results}

We now examine whether the temporally coarse-grained OU--Binomial baseline can
explain the aggregation-scale dependence documented above. The model is
estimated only at the monthly scale. Posterior sample paths
$\{p_t^{(s)}\}$ are then coarse-grained to longer horizons using the
survival-based aggregation relation introduced above. 
For each aggregation scale $k$, the model-implied count distribution is obtained
by evaluating the binomial model with the coarse-grained probabilities
$p_b^{(k,s)}$, rather than by re-estimating a separate latent process at each
horizon.

\begin{figure}[htbp]
    \centering
    \includegraphics[width=0.7\linewidth]{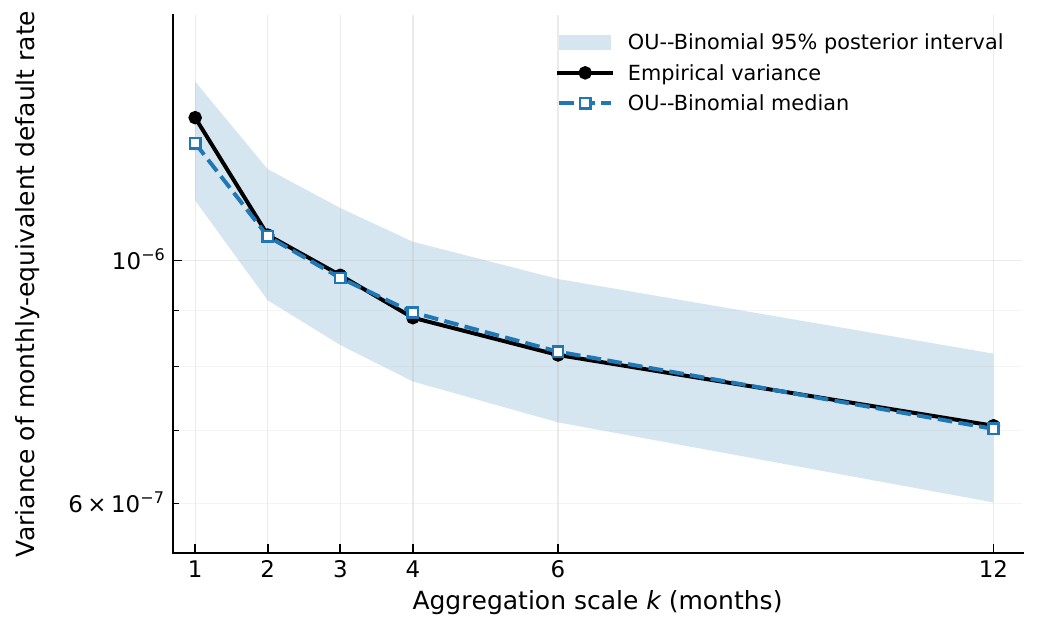}
    \caption{
Posterior predictive variance scaling of the monthly-equivalent default rate
under the coarse-grained OU--Binomial model. The solid line denotes the
posterior median, the shaded band denotes the 95\% posterior predictive
interval, and the points denote the empirical variance for
$k=1,2,3,4,6,12$ months.
}
\label{fig:ou_binomial_variance_scaling}
\end{figure}

Figure~\ref{fig:ou_binomial_variance_scaling} compares the empirical variance
of the monthly-equivalent default rate with the posterior predictive variance
implied by the coarse-grained OU--Binomial model. The empirical points are well
covered by the posterior predictive intervals over the whole range
$k=1,2,3,4,6,12$. In particular, the model reproduces the slow decay of the
variance with increasing aggregation scale. This result shows that persistent
monthly fluctuations of the latent default probability are already sufficient
to account for most of the long-horizon overdispersion observed in aggregated
default counts. In other words, much of the variance that may appear at longer
horizons as same-period default correlation is already generated by temporal
coarse-graining of the latent default-probability path, without introducing an
additional instantaneous-dependence parameter.

\begin{figure}[htbp]
    \centering
    \includegraphics[width=0.95\linewidth]{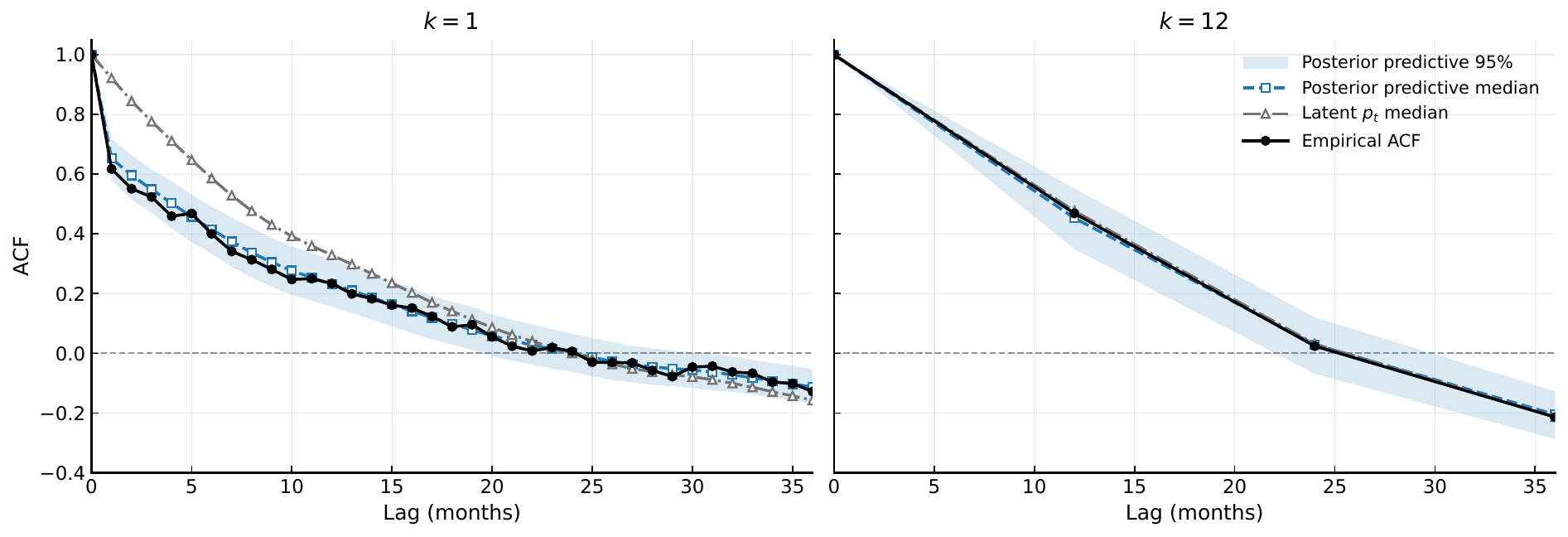}
    \caption{
Autocorrelation functions of the monthly-equivalent default rate under the
OU--Binomial model. The empirical ACF is compared with the posterior predictive
median and 95\% posterior predictive interval.
The latent ACF computed from the coarse-grained posterior paths is also shown.
}
\label{fig:ou_binomial_acf}
\end{figure}

We next examine temporal dependence more directly through the autocorrelation
function of the monthly-equivalent default rate. Figure~\ref{fig:ou_binomial_acf}
shows the empirical autocorrelation together with the posterior predictive
autocorrelation generated by the coarse-grained OU--Binomial paths for
$k=1$ and $k=12$. At the monthly scale, the posterior predictive median 
reproduces the empirical autocorrelation at short and intermediate lags and 
captures its gradual decay
toward zero. At the annual scale, used here as a representative long-horizon
case, the empirical autocorrelation is also broadly consistent with the
posterior predictive behavior generated by the same monthly latent
default-probability process. ACF diagnostics for all aggregation scales are
reported in Appendix~C.

This confirms that the aggregation-scale dependence of the variance shown in
Fig.~\ref{fig:ou_binomial_variance_scaling} is not merely a marginal
distributional effect. It is tied to temporal persistence in the latent
default-probability path. Thus, the same posterior dynamics that explain the
slow reduction of variance under temporal aggregation also explain the observed
autocorrelation structure.

\begin{figure}[htbp]
    \centering
    \includegraphics[width=0.9\linewidth]{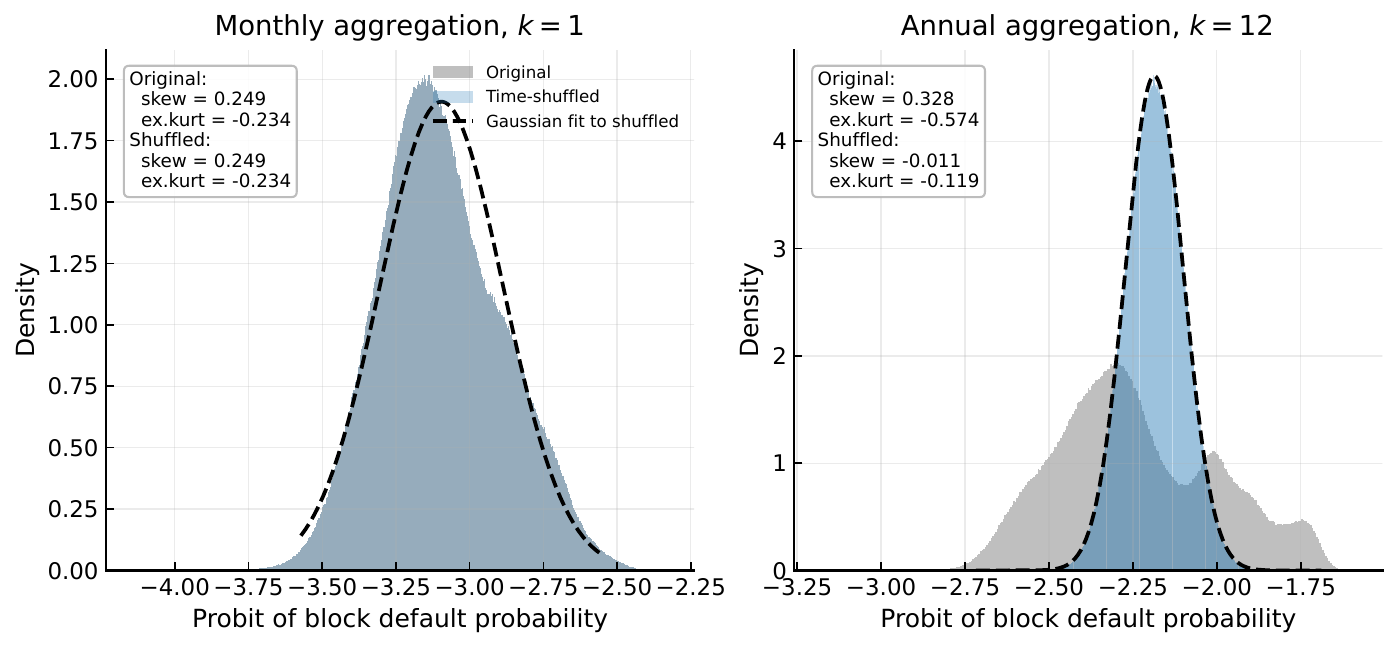}
\caption{
Effective mixing distributions on the probit scale induced by the posterior
monthly latent default-probability paths. For each aggregation scale $k$, the
distribution of $z_b^{(k,s)}=\Phi^{-1}(p_b^{(k,s)})$ is computed from the
Bayesian posterior samples of the coarse-grained default probabilities. The
original posterior paths are compared with time-shuffled paths that preserve
the one-month marginal posterior distribution of $p_t$ but destroy temporal
persistence. The dashed curve denotes a Gaussian fit to the shuffled benchmark.
}
\label{fig:effective_mixing_distribution}
\end{figure}

We then examine the effective mixing distribution induced by temporal
coarse-graining. For each posterior sample path, we compute the coarse-grained
probabilities $p_b^{(k,s)}$ and regard the collection
$\{p_b^{(k,s)}\}_{b,s}$ as a posterior-sample approximation to the effective
mixing distribution $G_k$. Thus, the $k$-month default-count distribution is a
binomial mixture whose mixing distribution is generated by coarse-graining the
monthly posterior latent path, rather than fitted independently at scale $k$.

To characterize this induced mixture, we transform the coarse-grained
probabilities to the probit scale,
$z_b^{(k,s)}=\Phi^{-1}(p_b^{(k,s)})$.
Figure~\ref{fig:effective_mixing_distribution} compares the resulting
distribution for the original posterior paths with a time-shuffled benchmark.
The shuffled paths preserve the one-month marginal distribution of $p_t$ but
destroy the temporal ordering of high- and low-default-probability states.
Thus, differences between the original and shuffled distributions isolate the
effect of temporal persistence in the monthly latent path.

At $k=1$, the two distributions coincide by construction. At the annual scale,
however, the original posterior paths generate an effective mixing distribution
that remains visibly different from the shuffled benchmark, while the shuffled
benchmark is closer to a Gaussian distribution on the probit scale. This shows
that the effective mixing distribution at scale $k$ is not determined only by
the one-month marginal distribution of $p_t$, but also by the temporal
organization of the monthly latent default-probability path. In this sense,
temporal coarse-graining generates a horizon-dependent effective mixing
distribution, which provides the distributional origin of apparent same-period
default correlation at aggregated observation scales.

As a scalar summary of this scale-dependent mixing distribution, we also
compute the effective default correlation
$\rho_D^{(k)}=\mathrm{Var}_{G_k}(p)/\{\bar p_k(1-\bar p_k)\}$,
where $\bar p_k=E_{G_k}[p]$.
In addition, we compute a probit-scale correlation index
$\chi_z^{(k)}=\mathrm{Var}_{G_k}(z)/\{1+\mathrm{Var}_{G_k}(z)\}$ with
$z=\Phi^{-1}(p)$.
For a Gaussian Vasicek mixing distribution this index coincides with the
asset-correlation parameter, whereas here it is used only as a compact summary
of the dispersion of the non-Gaussian effective mixing distribution. Both
quantities increase with the aggregation scale, as shown in
Fig.~\ref{figS:effective_default_probit_index_Gk}, showing that temporal
coarse-graining alone generates a scale-dependent effective default
correlation.

These results establish the OU--Binomial model as the baseline for the
remainder of the analysis. The monthly latent default-probability path, when
temporally coarse-grained, explains the aggregation-scale dependence of the
monthly-equivalent default-rate variance, reproduces a substantial part of the
autocorrelation structure, and induces a nontrivial long-horizon mixing
distribution. This induced mixture provides the reference structure against
which apparent same-period default correlation should be evaluated. Additional
instantaneous-dependence mechanisms should therefore be treated as residual
components beyond this coarse-grained latent-probability baseline.

Additional diagnostics for all aggregation scales are reported in
Appendix~\ref{app:ou_binomial_diagnostics}. These include the variance
decomposition of the coarse-grained OU--Binomial model for the monthly-equivalent
default rate into conditional binomial noise and latent default-probability
fluctuations, the full ACF comparison, moment diagnostics of the effective
mixing distribution, and the effective default-correlation diagnostics induced
by $G_k(p)$.

\section{Direct and renormalized fitting of residual covariance}

We now examine whether residual instantaneous dependence can be identified
beyond the temporally coarse-grained latent default-probability path. We compare
two diagnostic extensions of the OU--Binomial baseline: an OU--Vasicek model
with asset-correlation-type dependence and an OU--Davis--Lo model with
contagion-type dependence. The question is whether these same-period covariance
components improve predictive fit to aggregated default counts after the
monthly latent path has been mapped to longer horizons by temporal
coarse-graining.

\subsection{OU--Vasicek and OU--Davis--Lo extensions}

Both residual-dependence extensions retain the latent default-probability
process of the OU--Binomial baseline and add one parameter that generates
same-period dependence within each observation period.

The first extension is the OU--Vasicek model. In this model, defaults are driven
by a common Gaussian factor in addition to idiosyncratic shocks. Conditional on
the latent default probability $p_t$, the parameter $\rho_A$ controls the
strength of the common-factor component, or asset correlation. When
$\rho_A=0$, the model reduces to the conditional binomial model. A positive
$\rho_A$ generates same-period default correlation among obligors.

The second extension is the OU--Davis--Lo model. This model also has the
baseline default probability $p_t$, but adds a contagion parameter $q$. Defaults
first occur independently with probability $p_t$. Each independently defaulted
obligor can then trigger additional defaults of other obligors with probability
$q$. Thus, $q$ represents a cumulative contagion mechanism operating within the
same observation period. When $q=0$, the model again reduces to the conditional
binomial model.

Thus, the two extensions represent common-factor and contagion-induced forms of
residual instantaneous dependence, respectively. Detailed model definitions and
likelihood expressions are given in Appendix~\ref{app:models}.

\subsection{Direct fitting at each aggregation scale}

We first examine the direct fitting route. For each aggregation scale
$k=1,2,3,4,6,12$, the monthly data are converted into non-overlapping
$k$-month default-count observations, and the OU--Vasicek and OU--Davis--Lo
models are fitted separately at each scale. For the predictive comparison, the
OU--Binomial baseline is also fitted directly and separately at each
aggregation scale, using the same $k$-month observations.

In the OU--Vasicek model, the conditional default-count distribution is denoted
by
$$
P(L_t\mid n_t,p_t,\rho_A),
$$
where $p_t$ is the baseline default probability and $\rho_A$ is the asset
correlation parameter. The parameter $\rho_A$ generates same-period default
correlation within the observation period. In the OU--Davis--Lo model, the
corresponding distribution is written as
$$
P(L_t\mid n_t,p_t,q),
$$
where $q$ is the contagion parameter. Defaults first occur independently with
probability $p_t$, and the initially defaulted obligors can then induce
additional defaults of other obligors. Thus, $\rho_A$ and $q$ represent two
different forms of residual instantaneous dependence.

For residual-dependence models, the conditional variance contains an additional
same-period covariance term. Consider the monthly-equivalent default rate
$r_b^{(k)}=L_b^{(k)}/{k n_b^{(k)}}$ with model-implied default probability
$m_b^{(k)}$ and exposure $n_b^{(k)}$. If the conditional pairwise covariance
between default indicators is denoted by $c_b^{(k)}$, then the variance
components for $r_b^{(k)}$ are
$$
\frac{m_b^{(k)}{1-m_b^{(k)}}}{k^2 n_b^{(k)}},
\qquad
\frac{n_b^{(k)}-1}{k^2 n_b^{(k)}}c_b^{(k)},
\qquad
\mathrm{Var}_b\left(\frac{m_b^{(k)}}{k}\right).
$$
The first term is the binomial-noise contribution, the second is the residual
same-period covariance contribution, and the third is the contribution of
variation in the model-implied default probability.

In the OU--Vasicek model, $c_b^{(k)}$ is induced by the common Gaussian asset
factor, whereas in the OU--Davis--Lo model it is generated by the contagion
mechanism. The covariance share reported below is this residual covariance
contribution divided by the total posterior predictive variance of
$r_b^{(k)}$.

Predictive fit is evaluated by the expected log predictive density (elpd),
which is the sum of pointwise log predictive densities with a penalty for
effective model complexity, as estimated by WAIC. Larger elpd values indicate
better predictive fit. Because the number of non-overlapping blocks $B_k$
depends on the aggregation scale, we compare models using the per-block elpd
gain relative to the directly fitted OU--Binomial baseline,
$$
\Delta {\rm elpd}^{\rm block}_k
=\frac{{\rm elpd}^{\rm model}_k-
{\rm elpd}^{\rm OU\text{-}Binomial}_k}{B_k}.
$$
Positive values indicate better predictive fit than the OU--Binomial baseline,
whereas negative values indicate worse predictive fit. Equivalently, since
${\rm WAIC}=-2,{\rm elpd}$, a positive $\Delta {\rm elpd}^{\rm block}_k$
corresponds to a lower WAIC per block.

\begin{figure}[htbp]
\centering
\includegraphics[width=0.48\linewidth]{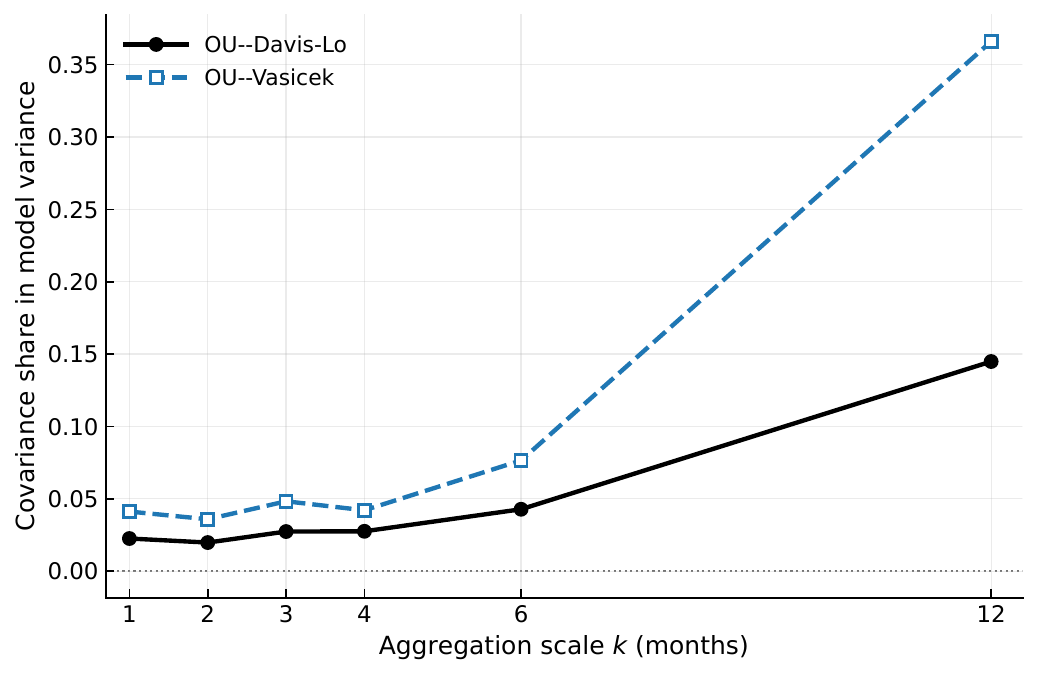}
\includegraphics[width=0.48\linewidth]{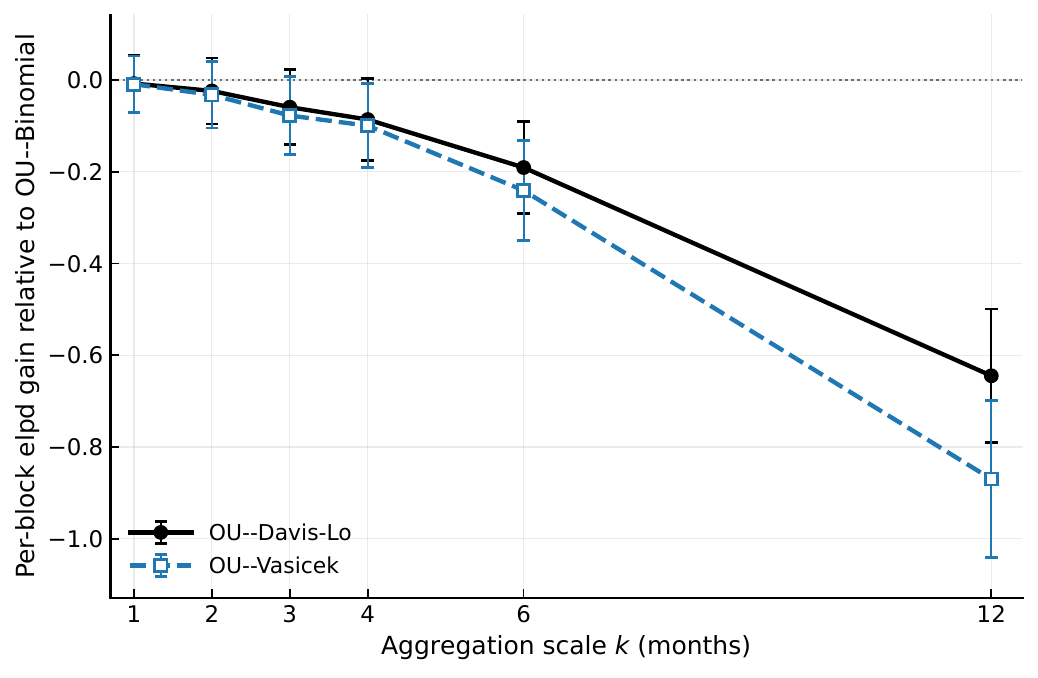}
\caption{
Direct-fitting diagnostics for the OU--Davis--Lo and OU--Vasicek
extensions.
(a) Share of the model variance assigned to the residual same-period
covariance component.
(b) Per-block elpd gain relative to the directly fitted OU--Binomial
baseline at the same aggregation scale. Error bars denote approximate standard
errors of the WAIC-based per-block elpd differences. Positive values indicate
improved predictive fit over the OU--Binomial baseline.
}
\label{fig:direct_fitting_diagnostics}
\end{figure}

Figure~\ref{fig:direct_fitting_diagnostics}(a) shows that the share of the
model variance assigned to the residual same-period covariance component
increases with the aggregation scale, especially in the OU--Vasicek model.
Taken alone, this behavior could be interpreted as evidence that same-period
default correlation becomes more important at longer horizons.

Figure~\ref{fig:direct_fitting_diagnostics}(b), however, shows that this
interpretation is not supported by predictive fit. The per-block elpd gain
relative to the directly fitted OU--Binomial baseline is negative for both
OU--Davis--Lo and OU--Vasicek, and its magnitude increases at longer aggregation
scales.

Thus, direct fitting can assign an increasing fraction of the model variance to
same-period covariance parameters even when these parameters do not improve
predictive density. The increasing covariance share should therefore be
interpreted as an apparent component, reflecting the absorption of unresolved
within-horizon latent default-probability fluctuations into 
instantaneous-dependence parameters.

\subsection{Renormalized fitting conditional on the coarse-grained latent path}

We next examine whether the instantaneous-dependence component remains
identifiable after the monthly latent default-probability path is temporally
coarse-grained. In contrast to direct fitting, the renormalized route does not
estimate a new latent process independently at each aggregation scale. Instead,
a model is first fitted at the monthly scale, and its posterior latent
default-probability paths are mapped to longer horizons by temporal
coarse-graining. The resulting coarse-grained posterior ensemble is then used as
the latent component of the long-horizon model. We refer to this procedure as
renormalized fitting in this operational sense.

At $k=1$, we fit the OU--Binomial, OU--Vasicek, and OU--Davis--Lo models to
the monthly default-count data. For each model, we extract posterior sample
paths of the monthly baseline default probability $p_t$. In the empirical
implementation below, we use $S=10^3$ posterior sample paths. For each posterior
sample path $s$, the monthly probabilities are coarse-grained by survival
aggregation,
$$
p_b^{(k,s)}=1-\prod_{j=0}^{k-1}{1-p_{bk+j}^{(s)}}.
$$
The collection ${p_b^{(k,s)}}_s$ defines the model-specific renormalized
latent default-probability path at aggregation scale $k$.

For $k\ge 2$, the residual dependence parameter is then estimated conditional
on this coarse-grained posterior ensemble. Specifically, for a residual
parameter $\theta_k$, where $\theta_k=q_k$ for the OU--Davis--Lo model and
$\theta_k=\rho_{A,k}$ for the OU--Vasicek model, we use the marginal likelihood
$$
P(L_b^{(k)}\mid n_b^{(k)},\theta_k)=
\frac{1}{S}
\sum_{s=1}^{S}
P(L_b^{(k)}\mid n_b^{(k)},p_b^{(k,s)},\theta_k),
$$
where $S$ is the number of posterior latent paths. The likelihood for the
$k$-month data is obtained by multiplying this marginal likelihood over
non-overlapping blocks. We then estimate $q_k$ or $\rho_{A,k}$ in a Bayesian
way using this path-marginalized likelihood. Thus, the uncertainty of the
coarse-grained latent default-probability path is retained when the residual
instantaneous dependence is re-estimated.

Thus, in the renormalized route, residual covariance is estimated conditional
on the coarse-grained latent default-probability path. The resulting covariance
component therefore represents residual instantaneous dependence beyond the
variation already captured by temporal coarse-graining.

\begin{figure}[htbp]
\centering
\includegraphics[width=0.48\linewidth]{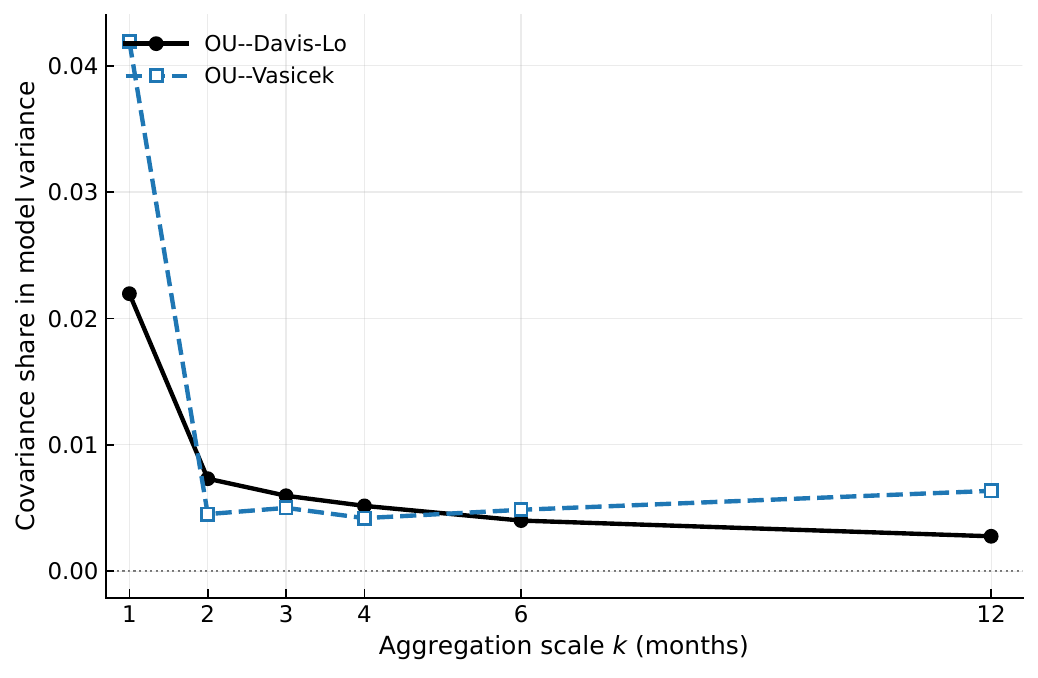}
\includegraphics[width=0.48\linewidth]{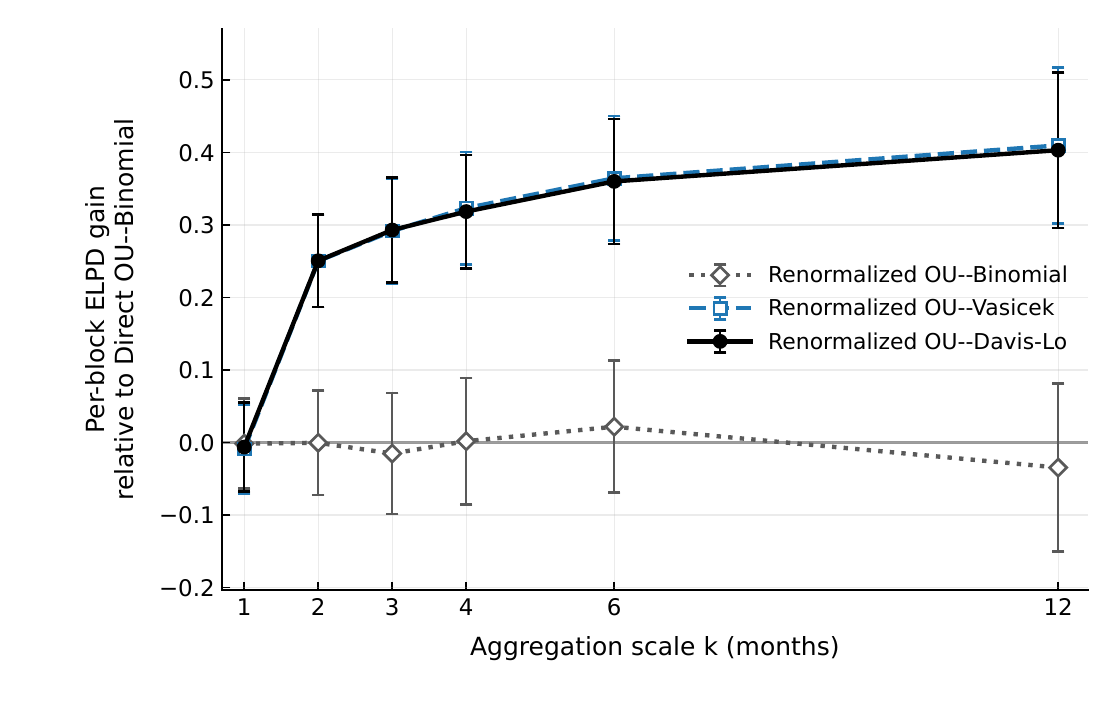}
\caption{
Renormalized-fitting diagnostics for the OU--Davis--Lo and OU--Vasicek
specifications.
(a) Share of the model variance assigned to the residual instantaneous
covariance component after the model-specific monthly latent
default-probability path is coarse-grained to scale $k$.
(b) Per-block elpd gain relative to the directly fitted OU--Binomial
baseline at the same aggregation scale. Error bars denote approximate standard
errors of the WAIC-based per-block elpd differences. The horizontal zero line
corresponds to the direct OU--Binomial baseline.
}
\label{fig:renormalized_fitting_diagnostics}
\end{figure}

Figure~\ref{fig:renormalized_fitting_diagnostics}(a) shows that, once the
model-specific monthly latent default-probability paths are coarse-grained, the
residual covariance share remains small for both the OU--Davis--Lo and
OU--Vasicek specifications. This contrasts with the direct-fitting result in
Fig.~\ref{fig:direct_fitting_diagnostics}(a), where the covariance share
increases with the aggregation scale.

Figure~\ref{fig:renormalized_fitting_diagnostics}(b) compares predictive fit
using the directly fitted OU--Binomial model as the common reference. The
renormalized OU--Binomial path is comparable to this direct baseline, although
it is obtained without re-estimating a new latent process at each aggregation
scale. The renormalized OU--Davis--Lo and OU--Vasicek specifications outperform
the direct OU--Binomial baseline.

The residual-parameter estimates reported in
Appendix~\ref{app:renormalized_fit} show that $\rho_{A,k}$ is sharply reduced
for $k\ge 2$, while $q_k$ remains small. Together with the small covariance
shares in Fig.~\ref{fig:renormalized_fitting_diagnostics}(a), this indicates
that the predictive improvement in
Fig.~\ref{fig:renormalized_fitting_diagnostics}(b) comes from avoiding an
over-allocation of long-horizon fluctuations to residual covariance parameters.
The coarse-grained latent default-probability path first accounts for the
dominant variance component, so the residual-dependence parameters are estimated
only as small correction terms that refine the shape of the predictive count
distribution.

Taken together, these results show that temporal coarse-graining regularizes
the attribution of variance across observation scales. It prevents unresolved
long-horizon fluctuations from being over-allocated to residual covariance
parameters and leaves only a small residual-dependence correction to the
predictive count distribution.

\section{Conclusion}

This paper showed that persistent dynamics of a latent default-probability path
can generate effective default correlation through temporal coarse-graining
across observation scales. Starting from monthly corporate default-count data,
we estimated a minimal OU--Binomial state-space model in which the latent
default probability evolves persistently on the probit scale. The posterior
paths of the monthly default probability were then coarse-grained to longer
horizons by survival-based temporal aggregation. This construction induces an
effective mixing distribution of the coarse-grained default probability, and
hence a long-horizon binomial mixture distribution, without fitting an
independent mixture model at each aggregation scale.

Empirically, the coarse-grained OU--Binomial path explains much of the
aggregation-scale dependence of default-rate variance and a substantial part of
the autocorrelation structure. The induced effective mixing distribution also
depends on the temporal ordering of high- and low-default-probability states,
not only on the one-month marginal distribution of default probabilities. These
results show that long-horizon overdispersion and apparent default correlation
can arise from persistent latent default-probability dynamics alone, even when
monthly defaults are conditionally independent given the latent path.

We then compared two routes for introducing residual instantaneous dependence.
In direct fitting, OU--Davis--Lo and OU--Vasicek models fitted independently at
each aggregation scale assign an increasing share of variance to residual
same-period covariance, especially at longer horizons. However, this additional
flexibility worsens the per-block elpd relative to the directly fitted
OU--Binomial baseline. Thus, direct fitting can over-allocate unresolved
within-horizon latent default-probability variation to contagion or
asset-correlation parameters, producing an apparent residual covariance
component that is not supported by improved predictive density.

In the renormalized route, residual dependence is first estimated at the
monthly scale, and the resulting posterior latent default-probability paths are
then coarse-grained to longer horizons. Residual-dependence parameters are then
estimated conditional on these coarse-grained latent paths. After this temporal
coarse-graining, the residual covariance share remains small, while the
per-block elpd improves relative to both the renormalized OU--Binomial specification and
the directly fitted OU--Binomial baseline. This improvement is achieved because
the coarse-grained latent path acts as a scale-consistent baseline that
suppresses the over-allocation of long-horizon fluctuations to residual
covariance parameters. The remaining small residual instantaneous-dependence
component can then improve the shape of the predictive count distribution
without becoming the dominant source of long-horizon variance.

These results clarify the origin of the identifiability problem. The difficulty
arises because fluctuations of a coarse-grained latent default-probability path
and residual same-period dependence can generate similar mixture distributions
of temporally aggregated default counts. Since these mechanisms can produce
overlapping count distributions after aggregation, direct fitting at each
aggregation scale can reassign unresolved long-horizon latent fluctuations to
contagion or asset-correlation parameters. By contrast, the renormalized route
first accounts for latent default-probability fluctuations through temporal
coarse-graining and then estimates residual dependence conditional on this
scale-consistent baseline. Thus, temporal coarse-graining regularizes the
separation between latent default-probability dynamics and residual same-period
dependence, improving identifiability across observation scales.

Several extensions remain open. Annual default counts may retain temporal
dependence even after coarse-graining, suggesting that conditional effective
mixing distributions could improve long-horizon prediction. Another natural
extension is a multi-sector formulation, in which sector-specific latent
default-probability paths and their cross-sector dependence are coarse-grained
jointly to study the emergence of long-horizon effective default correlation.

\clearpage

\appendix

\section{Construction of temporally aggregated monthly data}
\label{app:data_aggregation}

This appendix describes how the monthly default-count data are aggregated into
non-overlapping $k$-month blocks. The original monthly data consist of the
number of defaults $L_t$ and the corresponding number of obligors $n_t$ in
month $t$.

For each aggregation scale $k$, block $b$ consists of the months
$t=bk,bk+1,\ldots,bk+k-1$. The aggregated default count is defined as the sum of
monthly default counts within the block:
$$
L_b^{(k)}=\sum_{j=0}^{k-1} L_{bk+j}.
$$
For the exposure variable, we use the number of obligors at the beginning of
the block:
$$
n_b^{(k)}
=n_{bk}.
$$
The corresponding $k$-month default rate and monthly-equivalent default rate
are defined as
$$
\frac{L_b^{(k)}}{n_b^{(k)}},
\qquad
r_b^{(k)}=
\frac{1}{k}
\frac{L_b^{(k)}}{n_b^{(k)}}.
$$

This convention treats the block-level exposure as the initial portfolio size
at the beginning of the $k$-month horizon. It is consistent with the
survival-based coarse-graining of the model probability,
$$
p_b^{(k)}
=1-\prod_{j=0}^{k-1}
{1-p_{bk+j}},
$$
which represents the probability that an obligor present at the beginning of
the block defaults at least once during the $k$-month period.

There is an approximation in this construction because the number of obligors
can change within a block. Defaults observed in later months may include
defaults of obligors that entered after the first month of the block, whereas
the denominator $n_b^{(k)}=n_{bk}$ counts the beginning-of-block obligors. Thus,
the aggregated rate should be interpreted as an exposure-normalized block
default count rather than as an exact fixed-cohort default probability.

We expect the qualitative results to be insensitive to this exposure convention,
because default probabilities are small in the present data. The monthly default
probability is of order $10^{-3}$, and the annual default probability is of
order $10^{-2}$. Therefore, moderate changes in the obligor count within a block
produce only small changes in the expected number of defaults relative to the
scale of the long-horizon default-count fluctuations analyzed in the paper.

As a consistency check, we compare the annual series obtained from the
$k=12$ aggregation of the monthly S\&P data with the reported annual S\&P
series. The results are summarized in
Table~\ref{tab:k12_annual_consistency}. The two annual series are closely
aligned. The mean obligor counts are 3820.5 and 3817.9, respectively, and the
mean difference is only 2.6 obligors. The mean default counts are 60.15 and
59.63, with a mean difference of 0.53 defaults. The mean difference in annual
default rates is $-1.8\times 10^{-5}$, and its standard deviation is
$6.31\times 10^{-4}$.

These diagnostics indicate that the $k=12$ aggregation of the monthly data is
broadly consistent with the reported annual S\&P data. Therefore, although the
number of obligors may vary within a block, the beginning-of-block exposure
convention does not generate a substantial discrepancy at the annual scale. We
therefore expect the qualitative results to be insensitive to minor differences
in the precise exposure convention used in the temporal coarse-graining.

\begin{table}[htbp]
\centering
\caption{
Consistency check between the annual series obtained from the $k=12$
aggregation of the monthly S\&P data and the reported annual S\&P data.
The table reports summary statistics of obligor counts, default counts,
annual default rates, and their differences. Here
$\Delta n=n_{k=12}-n_{\rm S\&P}$,
$\Delta L=L_{k=12}-L_{\rm S\&P}$, and
$\Delta r=r_{k=12}-r_{\rm S\&P}$.
}
\label{tab:k12_annual_consistency}
\begin{tabular}{lrrrrrrrrr}
\hline
Statistic
& $n_{k=12}$ & $n_{\rm S\&P}$
& $L_{k=12}$ & $L_{\rm S\&P}$
& $r_{k=12}$ & $r_{\rm S\&P}$
& $\Delta n$ & $\Delta L$ & $\Delta r$ \\
\hline
Count & 40.000 & 40.000 & 40.000 & 40.000 & 40.000 & 40.000 & 40.000 & 40.000 & 40.000 \\
Mean  & 3820.475 & 3817.875 & 60.150 & 59.625 & 0.014962 & 0.014980 & 2.600 & 0.525 & $-1.8\times10^{-5}$ \\
Std.  & 1689.913 & 1681.782 & 52.719 & 51.603 & 0.010089 & 0.010069 & 10.705 & 2.846 & $6.31\times10^{-4}$ \\
Min.  & 1342.000 & 1354.000 & 2.000 & 2.000 & 0.001490 & 0.001477 & -12.000 & -4.000 & $-1.496\times10^{-3}$ \\
25\%  & 2058.500 & 2063.250 & 19.500 & 20.000 & 0.008301 & 0.008088 & -1.500 & -1.000 & $-4.40\times10^{-4}$ \\
50\%  & 4290.000 & 4289.000 & 45.000 & 42.000 & 0.011808 & 0.011860 & 1.000 & 0.000 & $4.9\times10^{-5}$ \\
75\%  & 5130.750 & 5128.250 & 87.750 & 88.250 & 0.018260 & 0.018487 & 4.000 & 1.250 & $2.71\times10^{-4}$ \\
Max.  & 6284.000 & 6263.000 & 215.000 & 214.000 & 0.042532 & 0.042351 & 40.000 & 9.000 & $1.379\times10^{-3}$ \\
\hline
\end{tabular}
\end{table}

\clearpage

\section{Model definitions}
\label{app:models}

This appendix summarizes the one-period default-count distributions used as
instantaneous-dependence extensions of the OU--Binomial baseline. The
OU--Binomial state-space model is defined in the main text. Here we only define
the Vasicek and Davis--Lo conditional default-count models used in
Section~III.

\subsection{Vasicek default-count distribution}

In the Vasicek specification, default dependence is generated by a common
Gaussian factor. The baseline default probability $p_t$ is inherited from the
latent default-probability path of the OU--Binomial state-space model. As in
the main text, we write
$$
y_t=\Phi^{-1}(p_t),
\qquad
p_t=\Phi(y_t),
$$
where $\Phi$ denotes the standard normal cumulative distribution function.
For obligor $i$ in period $t$, define the latent variable
$$
Y_{it}
=
\sqrt{\rho_A}\,F_t
+
\sqrt{1-\rho_A}\,\epsilon_{it},
$$
where $F_t\sim N(0,1)$ is a common factor and
$\epsilon_{it}\sim N(0,1)$ are independent idiosyncratic shocks. Obligor $i$
defaults when
$$
Y_{it}\leq y_t .
$$
The parameter $\rho_A$ controls the strength of the common-factor component.
When $\rho_A=0$, defaults are conditionally independent given $p_t$.

Conditional on $F_t=f$, the default probability is
$$
p_t(f)
=
\Phi\left(
\frac{y_t-\sqrt{\rho_A}\,f}{\sqrt{1-\rho_A}}
\right).
$$
Therefore,
$$
L_t\mid n_t,p_t,\rho_A,F_t=f
\sim
\mathrm{Binomial}\{n_t,p_t(f)\}.
$$
After integrating out the common factor, the one-period default-count
distribution is
$$
P(L_t=\ell\mid n_t,p_t,\rho_A)
=
\int
\binom{n_t}{\ell}
p_t(f)^\ell
\{1-p_t(f)\}^{n_t-\ell}
\phi(f)\,df ,
$$
where $\phi$ is the standard normal density. Thus, the Vasicek model is a
continuous binomial mixture induced by the common Gaussian factor.

\subsection{Davis--Lo default-count distribution}

In the Davis--Lo specification, default dependence is generated by a cumulative
contagion mechanism. As in the main text, the baseline probability $p_t$ is
generated by the latent default-probability path of the OU--Binomial
state-space model. Let $X_{it}$ denote the idiosyncratic default indicator of
obligor $i$ in period $t$, with
$$
X_{it}\sim \mathrm{Bernoulli}(p_t).
$$
For each ordered pair $i\neq j$, let $Y_{ijt}$ denote the contagion indicator
from obligor $j$ to obligor $i$, with
$$
Y_{ijt}\sim \mathrm{Bernoulli}(q).
$$
All $X_{it}$ and $Y_{ijt}$ are assumed independent conditional on $p_t$ and
$q$. The final default indicator is
$$
Z_{it}
=
X_{it}
+
(1-X_{it})
\left[
1-
\prod_{j\neq i}
(1-Y_{ijt}X_{jt})
\right].
$$
Thus, defaults first occur independently with probability $p_t$, and each
initially defaulted obligor can trigger additional defaults of other obligors
with probability $q$.

Let
$$
K_t=\sum_{i=1}^{n_t}X_{it}
$$
be the number of initial idiosyncratic defaults. Then
$$
K_t\sim \mathrm{Binomial}(n_t,p_t).
$$
Conditional on $K_t=h$, each surviving obligor is infected with probability
$$
r_h=1-(1-q)^h.
$$
Therefore,
$$
L_t\mid K_t=h,n_t,p_t,q
\sim
h+\mathrm{Binomial}(n_t-h,r_h).
$$
Equivalently, for $\ell=0,1,\ldots,n_t$, the one-period default-count
distribution is
$$
P(L_t=\ell\mid n_t,p_t,q)
=
\sum_{h=0}^{\ell}
\binom{n_t}{h}
p_t^h(1-p_t)^{n_t-h}
\binom{n_t-h}{\ell-h}
r_h^{\ell-h}
(1-r_h)^{n_t-\ell}.
$$
The parameter $q$ controls the strength of cumulative contagion within the same
observation period. When $q=0$, the model reduces to the conditional binomial
model with default probability $p_t$.

The same-period covariance contributions discussed in Section~III are evaluated
from the posterior predictive distributions of these one-period extensions.

\subsection{Coarse-grained mixing distribution of the OU--Binomial baseline}

The preceding subsections defined the one-period default-count distributions
used as residual instantaneous-dependence extensions. We now give an analytical
characterization of the effective mixing distribution induced by temporal
coarse-graining of the OU--Binomial baseline. This characterization clarifies
that the posterior-sample construction used in the main text is a Monte Carlo
approximation to a finite-dimensional Gaussian integral.

In the OU--Binomial baseline, the monthly latent default probability is written
as
\[
p_t=\Phi(y_t),
\]
where the probit-scale latent state follows the stationary AR(1) process
\[
y_t-\mu=
\phi(y_{t-1}-\mu)+\varepsilon_t,
\qquad
\varepsilon_t\sim N(0,\sigma_\varepsilon^2).
\]
Equivalently, the process may be viewed as the monthly discrete-time
representation of an Ornstein--Uhlenbeck process. Under stationarity, the
variance of $y_t$ is
\[
\sigma_y^2=\frac{\sigma_\varepsilon^2}{1-\phi^2}.
\]
Since $p_t=\Phi(y_t)$ and, under stationarity,
$y_t\sim N(\mu,\sigma_y^2)$, a first-order delta-method approximation gives
\[
p_t=\Phi(y_t)\simeq\Phi(\mu)+\varphi(\mu)(y_t-\mu),
\]
where \(\varphi\) denotes the standard normal density. Hence,
\[
p_t\approx N\left(\Phi(\mu),\varphi(\mu)^2\sigma_y^2\right).
\]
This approximation is local and is used only to interpret the scale of
monthly latent-probability fluctuations.

For a block of $k$ consecutive months, let
\[
\mathbf{y}^{(k)}=(y_1,\ldots,y_k)^\top
\]
denote a stationary segment of the latent AR(1) process. Then
\[
\mathbf{y}^{(k)}\sim N(\mu\mathbf{1},\Sigma_k),
\]
where
\[
(\Sigma_k)_{ij}=\sigma_y^2\phi^{|i-j|}=
\frac{\sigma_\varepsilon^2}{1-\phi^2}\phi^{|i-j|}.
\]

Define the survival-aggregation map
\[
T_k(\mathbf{y}^{(k)})=T_k(y_1,\ldots,y_k)=1-\prod_{j=1}^{k}\{1-\Phi(y_j)\}.
\]
The $k$-month coarse-grained default probability is then 
$p^{(k)}=T_k(\mathbf{y}^{(k)})$.
Accordingly, the effective mixing distribution at aggregation scale $k$ is
the distribution $G_k$ of $p^{(k)}$. Equivalently, for any measurable set
$A\subset[0,1]$,
\[
G_k(A)=
\int_{\mathbb{R}^k}\mathbf{1}\{T_k(\mathbf{y})\in A\}
\varphi_k(\mathbf{y};\mu\mathbf{1},\Sigma_k)
\,d\mathbf{y},
\]
where $\varphi_k(\cdot;\mu\mathbf{1},\Sigma_k)$ denotes the $k$-variate normal
density.

Conditional on the coarse-grained default probability $p^{(k)}=p$, the
$k$-month default count follows
\[
L^{(k)}\mid p^{(k)}=p,n^{(k)}\sim \mathrm{Binomial}(n^{(k)},p).
\]
Therefore, the unconditional $k$-month default-count distribution is the
binomial mixture
\[
P(L^{(k)}=\ell)=\int_0^1 \binom{n^{(k)}}{\ell}
p^\ell(1-p)^{n^{(k)}-\ell}
\,dG_k(p).
\]

In general, $G_k$ does not have an elementary closed-form density, because the
map $T_k$ contains Gaussian cumulative distribution functions and a product of
monthly survival probabilities. Nevertheless, $G_k$ is analytically well
defined as the distribution induced by applying the survival-aggregation map to
a finite-dimensional Gaussian AR(1) vector. The posterior effective mixing
distributions shown in the main text are posterior Monte Carlo approximations
to this analytically defined object, with parameter uncertainty and latent-path
uncertainty retained through the Bayesian posterior.

A useful first-order approximation is obtained by applying the delta method to
the survival-aggregation map $T_k$. At the stationary mean,
\[
T_k(\mu\mathbf{1})=1-\{1-\Phi(\mu)\}^k .
\]
Moreover,
\[
\left.\frac{\partial T_k}{\partial y_i}\right|_{\mathbf{y}=\mu\mathbf{1}}=\varphi(\mu)\{1-\Phi(\mu)\}^{k-1},
\]
where \(\varphi\) denotes the standard normal density. Therefore,
\[
p^{(k)}\approx N\left(
1-\{1-\Phi(\mu)\}^k,\,\varphi(\mu)^2\{1-\Phi(\mu)\}^{2(k-1)}\mathbf{1}^\top\Sigma_k\mathbf{1}\right).
\]
Since
\[
\mathbf{1}^\top\Sigma_k\mathbf{1}=\sigma_y^2\left[k+2\sum_{h=1}^{k-1}(k-h)\phi^h\right],
\]
we obtain
\[
\mathrm{Var}(p^{(k)})\approx \varphi(\mu)^2\{1-\Phi(\mu)\}^{2(k-1)}\sigma_y^2
\left[k+2\sum_{h=1}^{k-1}(k-h)\phi^h\right].
\]
For the monthly-equivalent default probability \(p^{(k)}/k\),
\[
\mathrm{Var}\left(\frac{p^{(k)}}{k}\right) \approx
\frac{\varphi(\mu)^2\{1-\Phi(\mu)\}^{2(k-1)}}{k^2}\sigma_y^2
\left[k+2\sum_{h=1}^{k-1}(k-h)\phi^h\right].
\]
This approximation shows explicitly how positive AR(1) persistence inflates
the variance of the coarse-grained monthly-equivalent default probability
relative to the independent benchmark.

This approximation is used only to interpret the variance scaling; it is not
used to replace the exact effective mixing distribution \(G_k\), which remains
non-Gaussian in general.


\clearpage

\section{Additional diagnostics for the coarse-grained OU--Binomial model}
\label{app:ou_binomial_diagnostics}

\subsection{Bayesian estimation of the monthly OU--Binomial model}

The monthly OU--Binomial model was estimated by Bayesian inference using the
monthly S\&P default-count series from January 1981 to September 2021
($T=489$). The observation equation is
$$
L_t \mid p_t,n_t \sim {\rm Binomial}(n_t,p_t),
$$
and the latent default probability is represented on the probit scale as
$$
y_t=\Phi^{-1}(p_t),
\qquad
p_t=\Phi(y_t).
$$
To describe persistent fluctuations of the latent default probability, the
latent state was modeled as a stationary AR(1), or discrete-time OU, process,
$$
y_t-\mu=\phi(y_{t-1}-\mu)+\epsilon_t,
\qquad
\epsilon_t\sim N(0,\sigma_\epsilon^2).
$$
Equivalently, writing $x_t=y_t-\mu$, the centered state satisfies
$$
x_t=\phi x_{t-1}+\eta_t,
\qquad
\eta_t\sim N(0,\sigma_\eta^2),
$$
with $\sigma_\eta=\sigma_\epsilon$.
The initial centered state was drawn from the stationary distribution,
$$
x_0\sim N\left(0,\frac{\sigma_\eta^2}{1-\phi^2}\right).
$$
The stationary standard deviation of the latent probit-scale default-probability state is
$$
\sigma_x=\frac{\sigma_\eta}{\sqrt{1-\phi^2}}.
$$

The prior distributions were chosen to favor a persistent but flexible monthly
latent default-probability process:
$$
\mu\sim N(\mu_0,1^2),\qquad
\phi\sim {\rm Beta}(20,2),\qquad
\sigma_\eta\sim {\rm HalfNormal}(0.20),
$$
where $\mu_0$ was initialized from a preliminary empirical estimate of the
monthly default level. Posterior sampling was performed using the NUTS sampler
with four chains, 5000 tuning iterations, and 2000 retained draws per chain,
giving 8000 posterior draws in total.

Table~\ref{tab:ou_binom_posterior_summary} reports the posterior summary of the
main parameters. The posterior median of the persistence parameter is
approximately $\phi=0.942$, corresponding to a half-life of about 13 months.
This confirms that the latent default-probability state is highly persistent at
the monthly scale. The posterior standard deviation of the stationary latent
state is about $0.23$ on the probit scale.

\begin{table}[t]
\centering
\caption{
Posterior summary of the monthly OU--Binomial state-space model.
The half-life is measured in months and is defined as
$\log(0.5)/\log(\phi)$.
}
\label{tab:ou_binom_posterior_summary}
\begin{tabular}{lrrrrrr}
\hline
Parameter & Mean & SD & 2.5\% HDI & 97.5\% HDI & ESS & $\hat R$ \\
\hline
$\mu$          & $-3.113$ & $0.059$ & $-3.237$ & $-3.010$ & $1192$  & $1.01$ \\
$\phi$        & $0.942$  & $0.018$ & $0.906$  & $0.977$  & $3374$  & $1.00$ \\
$\sigma_\eta$ & $0.074$  & $0.007$ & $0.060$  & $0.088$  & $2316$  & $1.01$ \\
$\sigma_x$    & $0.227$  & $0.038$ & $0.168$  & $0.301$  & $3128$  & $1.00$ \\
Half-life     & $13.151$ & $5.962$ & $5.709$  & $23.479$ & $3374$  & $1.00$ \\
$\kappa=-\log\phi$ & $0.060$ & $0.020$ & $0.023$ & $0.099$ & $3374$ & $1.00$ \\
\hline
\end{tabular}
\end{table}

The convergence diagnostics are also reported in
Table~\ref{tab:ou_binom_posterior_summary}. The rank-normalized $\hat R$
values are close to unity for all scalar parameters. The effective sample size
is smaller for $\mu$ than for the other parameters, reflecting posterior
dependence between the long-run level and the latent path, but it is still
sufficient for the coarse-graining analysis. Across the 489 latent monthly
probabilities $p_t$, the posterior diagnostics were stable, with a mean bulk
effective sample size of about $1.06\times 10^4$ and a maximum $\hat R$ of
approximately $1.002$.

\subsection{Variance decomposition of the coarse-grained OU--Binomial model}

We first examine how the variance of the coarse-grained OU--Binomial model is
decomposed into conditional binomial noise and fluctuations of the latent
default-probability path. For each aggregation scale $k$, we decompose the
posterior predictive variance of the monthly-equivalent default rate
$r_b^{(k)}=L_b^{(k)}/\{k n_b^{(k)}\}$ into the binomial-noise contribution and
the latent default-probability contribution. Each component is normalized by
the empirical variance of the monthly-equivalent default rate at the same
aggregation scale.

\begin{figure}[htbp]
    \centering
    \includegraphics[width=0.72\linewidth]{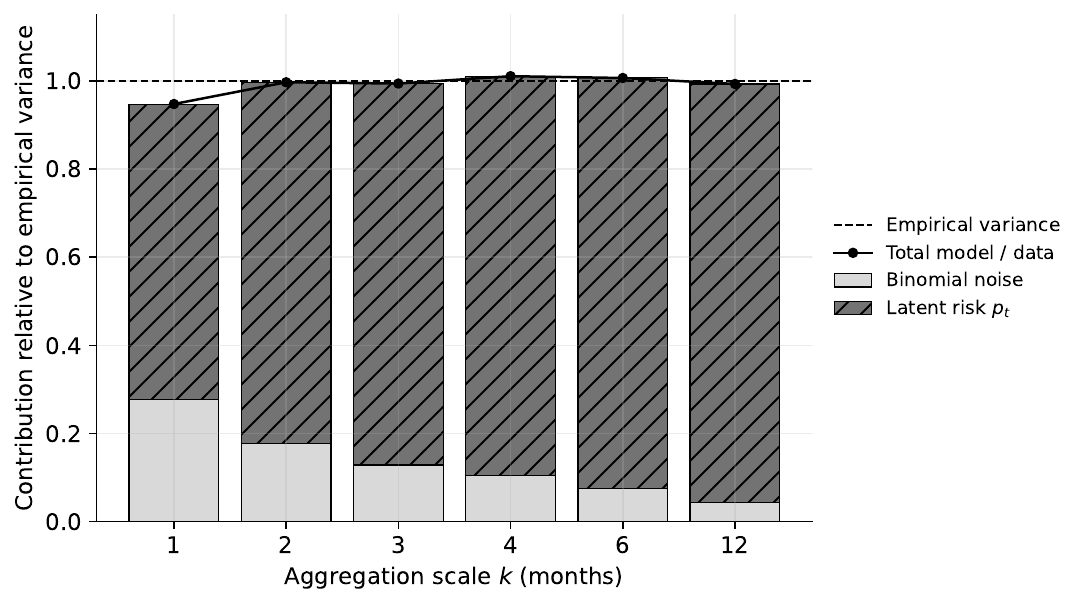}
\caption{
Variance decomposition of the coarse-grained OU--Binomial model for the
monthly-equivalent default rate. The posterior median model variance is
decomposed into conditional binomial noise and the contribution of the
coarse-grained latent default-probability path. The dashed line indicates the
empirical variance, and the solid line with markers shows the total model
variance normalized by the empirical variance.
}
\label{fig:app_ou_binom_variance_decomposition}
\end{figure}

Figure~\ref{fig:app_ou_binom_variance_decomposition} shows the resulting
decomposition. The total model variance is close to the empirical variance for
all aggregation scales. At $k=1$, conditional binomial noise accounts for about
28\% of the empirical monthly-equivalent variance, while latent
default-probability fluctuations account for about 67\%. As $k$ increases, the
binomial-noise contribution rapidly decreases, reaching about 4\% at $k=12$.
In contrast, the latent default-probability contribution becomes dominant and
accounts for about 95\% of the empirical variance at the annual horizon.

This decomposition supports the interpretation used in the main text. The slow
decay of the observed monthly-equivalent default-rate variance with aggregation
scale is mainly explained by persistent fluctuations in the latent
default-probability path rather than by conditional binomial noise.

\subsection{Autocorrelation diagnostics}

\begin{figure}[htbp]
    \centering
    \includegraphics[width=0.8\linewidth]{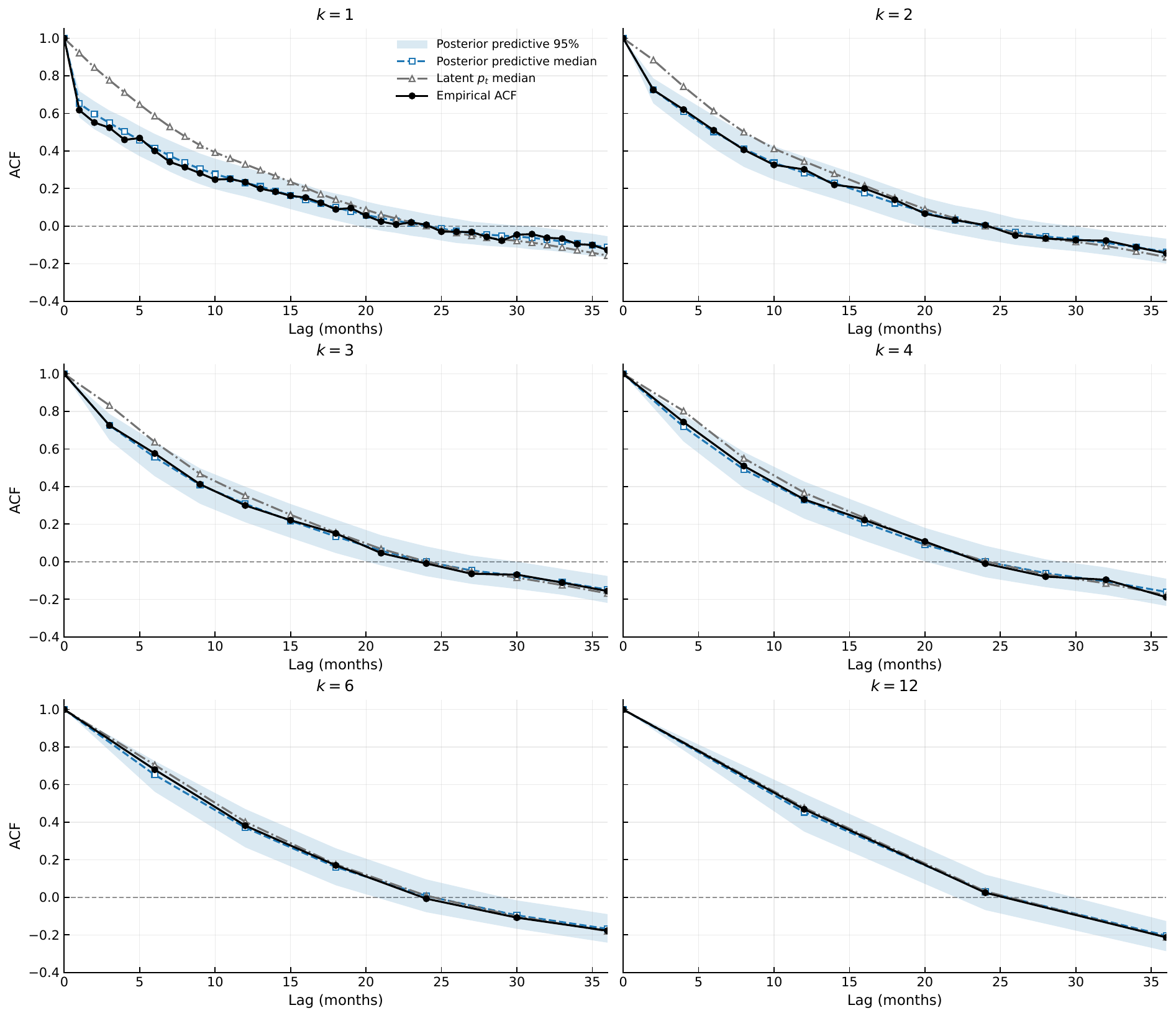}
    \caption{
Autocorrelation functions of the monthly-equivalent default rate for all
aggregation scales under the coarse-grained OU--Binomial model.
The empirical ACF is compared with the posterior predictive median and
95\% posterior predictive interval. The ACF of the latent coarse-grained
default-probability path is also shown. The same monthly OU--Binomial posterior
paths reproduce the persistence structure over $k=1,2,3,4,6,12$.
}
    \label{fig:app_ou_binom_acf_all_k}
\end{figure}

Figure~\ref{fig:app_ou_binom_acf_all_k} shows that the coarse-grained
OU--Binomial posterior paths reproduce the empirical ACF over all aggregation
scales $k=1,2,3,4,6,12$, confirming that the variance scaling is accompanied by
a consistent representation of temporal persistence in the latent
default-probability path.

\subsection{Effective mixing distributions}

\begin{figure}[htbp]
    \centering
    \includegraphics[width=0.8\linewidth]{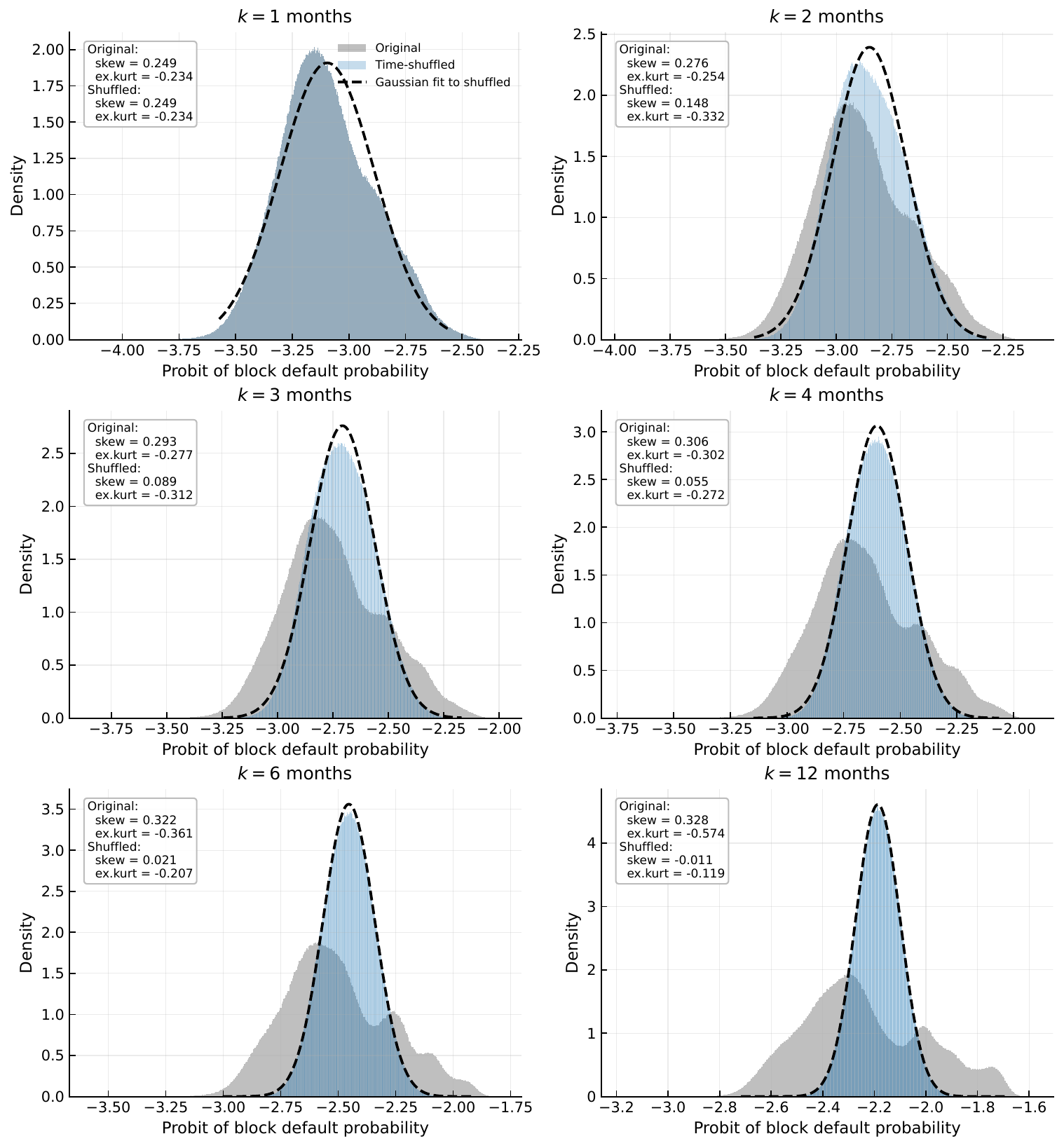}
    \caption{
    Effective mixing distributions on the probit scale for all aggregation
    scales. For each $k$, the distribution of
    $z_b^{(k,s)}=\Phi^{-1}(p_b^{(k,s)})$ is shown for the original posterior
    paths and for time-shuffled paths. The shuffled paths preserve the
    one-month marginal distribution of $p_t$ but destroy temporal ordering.
    At $k=1$, the original and shuffled distributions coincide by construction.
    As $k$ increases, the original distribution increasingly differs from the
    shuffled benchmark, showing that the effective long-horizon mixing
    distribution depends not only on the marginal distribution of monthly
    latent default-probability states but also on their temporal persistence.
    }
    \label{fig:app_probit_distribution_all_k}
\end{figure}

\begin{figure}[t]
    \centering
    \includegraphics[width=0.75\linewidth]{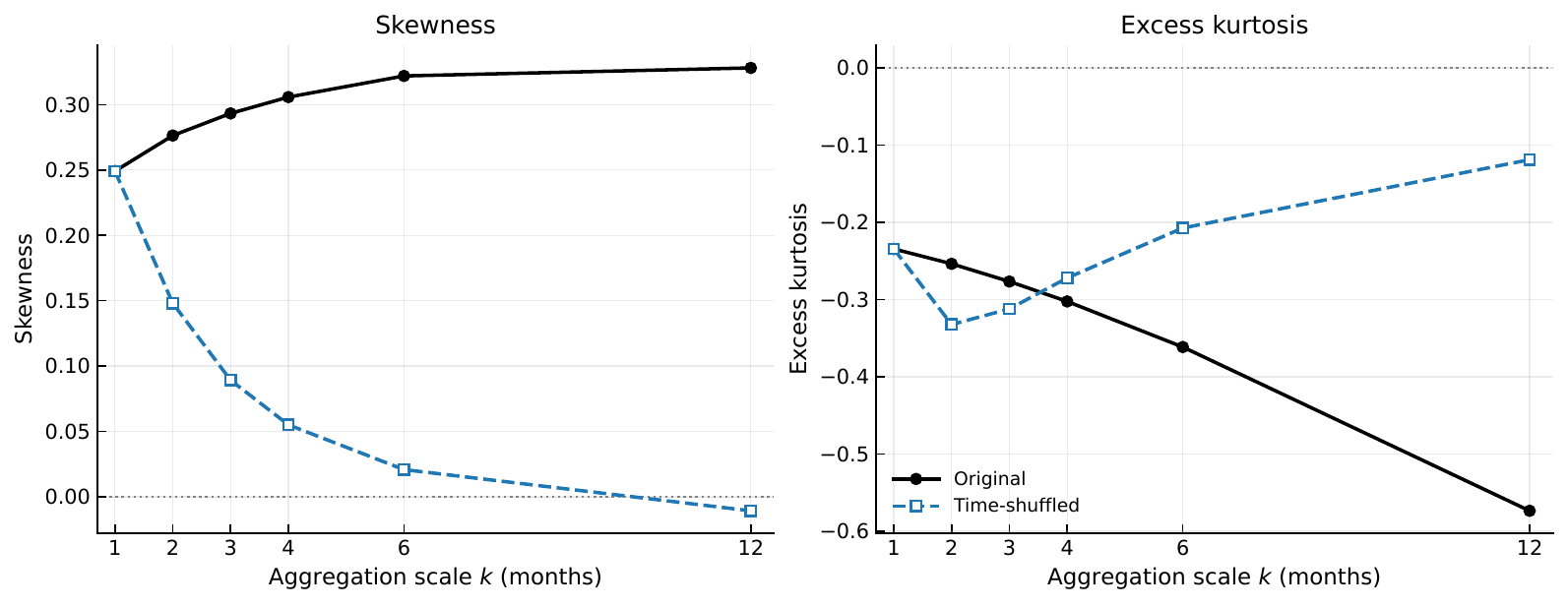}
    \caption{
    Skewness and excess kurtosis of the effective mixing distribution on the
    probit scale as functions of the aggregation scale $k$. The original
    temporally ordered paths retain positive skewness as $k$ increases, whereas
    the time-shuffled benchmark becomes nearly symmetric. The excess kurtosis
    also evolves differently between the original and shuffled paths. These
    moment diagnostics provide a compact summary of the temporal-ordering
    effect shown in Fig.~\ref{fig:app_probit_distribution_all_k}.
    }
    \label{fig:app_skewness_kurtosis}
\end{figure}

Figures~\ref{fig:app_probit_distribution_all_k} and
\ref{fig:app_skewness_kurtosis} provide additional diagnostics of the effective
mixing distribution. The original posterior paths and the time-shuffled
benchmark differ increasingly with the aggregation scale, indicating that the
long-horizon distribution is shaped by the temporal ordering and persistence of
the monthly latent default-probability path, not only by its marginal
distribution.

\begin{figure}[htbp]
    \centering
    \includegraphics[width=0.95\linewidth]{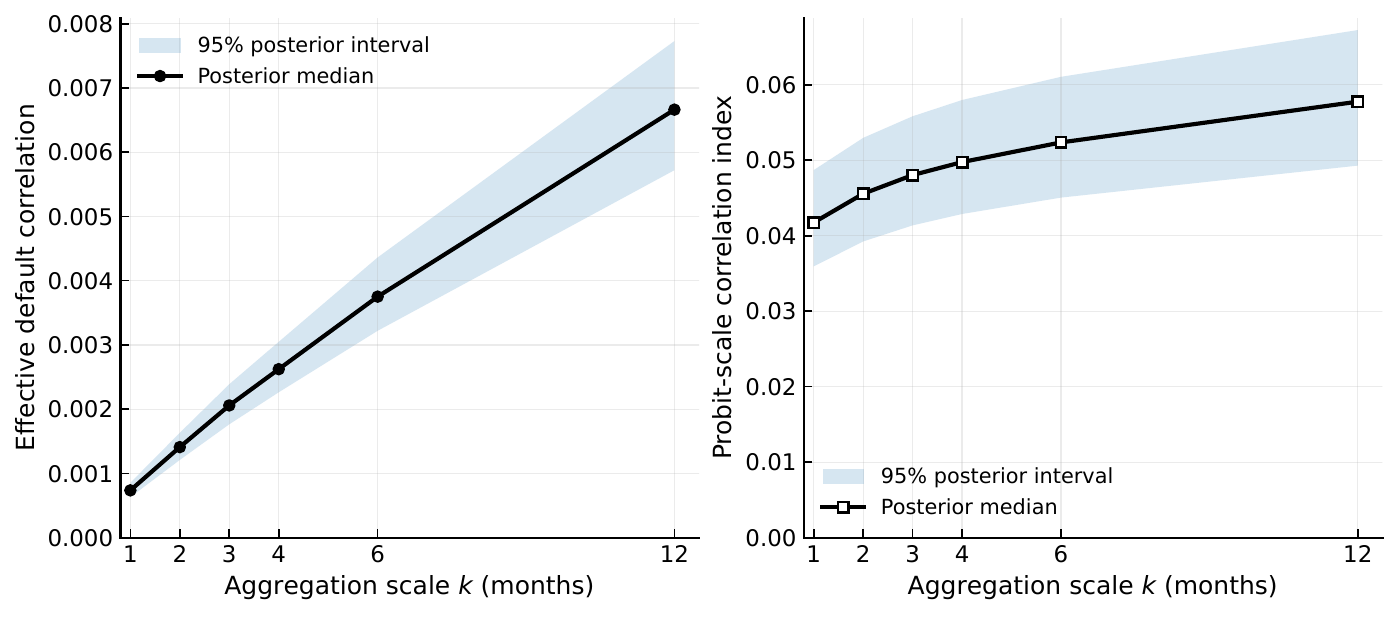}
    \caption{
    Effective default correlation and probit-scale correlation index
    induced by the effective mixing distribution $G_k(p)$.
    For each aggregation scale $k$, the effective default correlation is
    computed as
    $\rho_D^{(k)}=\mathrm{Var}_{G_k}(p)/
    \{\bar p_k(1-\bar p_k)\}$,
    where $\bar p_k=E_{G_k}[p]$.
    The probit-scale index is computed from the variance of the
    probit-transformed mixing distribution,
    $z=\Phi^{-1}(p)$, as
    $\chi_z^{(k)}=\mathrm{Var}_{G_k}(z)/
    \{1+\mathrm{Var}_{G_k}(z)\}$.
    For a Gaussian Vasicek mixing distribution this index coincides with
    the asset-correlation parameter, but here it is used only as a
    scale-dependent summary of the dispersion of the non-Gaussian
    effective mixing distribution.
    The increase with $k$ shows that temporal coarse-graining of persistent
    latent default-probability dynamics generates a scale-dependent effective
    default correlation, even in the absence of instantaneous default
    dependence.
    }
    \label{figS:effective_default_probit_index_Gk}
\end{figure}

\clearpage

\section{Direct fitting details}
\label{app:direct_fit}

This appendix summarizes the direct-fitting procedure used in
Sec.~III~B. The analysis uses the S\&P monthly default-count series for
the ALL sector. For each aggregation scale $k=1,2,3,4,6,12$, the monthly
observations are converted into non-overlapping $k$-month blocks. The
numbers of blocks are
$B_k=489,244,163,122,81,40$, respectively.

In the direct-fitting route, the OU--Binomial, OU--Davis--Lo, and
OU--Vasicek models are fitted independently at each aggregation scale.
Thus, both the latent default-probability path and, when present, the residual
dependence parameter are re-estimated separately for each $k$. This procedure
is therefore different from the renormalized fitting route, where the monthly
posterior latent paths are first estimated and then coarse-grained to longer
horizons.

For all direct fits, the baseline default probability is represented on
the probit scale,
$$
p_b^{(k)}=\Phi(y_b^{(k)}),
$$
and the latent state follows the same stationary OU/AR(1) specification as
in Appendix~\ref{app:ou_binomial_diagnostics}, but with scale-dependent
parameters $(\mu_k,\phi_k,\sigma_{\eta,k})$. The prior distributions and
NUTS sampling settings also follow the monthly Bayesian specification,
except that the OU persistence prior is
$$
\phi_k \sim {\rm Beta}(30,2).
$$
For the residual-dependence extensions, we use
$$
q_k\sim {\rm Beta}(2,50)
$$
for the OU--Davis--Lo model and
$$
\rho_{A,k}\sim {\rm Beta}(2,8)
$$
for the OU--Vasicek model. The Vasicek likelihood is evaluated by
Gauss--Hermite quadrature with 100 nodes.

The same variance-decomposition formula is used for both the direct-fitting
and renormalized-fitting diagnostics.
For each model and aggregation scale, the posterior predictive variance of the
monthly-equivalent default rate
$$
r_b^{(k)}=
\frac{L_b^{(k)}}{k n_b^{(k)}}
$$
is decomposed into conditional binomial noise, variation of the model-implied
default probability, and residual instantaneous covariance. 
For a block-level
default probability $m_b^{(k)}$ and pairwise same-period covariance
$c_b^{(k)}$, the three components are
$$
\frac{m_b^{(k)}\{1-m_b^{(k)}\}}{k^2 n_b^{(k)}},
\qquad
\frac{n_b^{(k)}-1}{k^2 n_b^{(k)}}c_b^{(k)},
\qquad
{\rm Var}_b
\left(
\frac{m_b^{(k)}}{k}
\right).
$$
The second term is absent for the OU--Binomial model. In the OU--Davis--Lo
model, $c_b^{(k)}$ is induced by $q_k$, while in the OU--Vasicek model it is
induced by $\rho_{A,k}$. The covariance share reported in the main text is the
residual covariance component divided by the total model variance.

For each fitted model, WAIC is computed from the pointwise log-likelihood
over the non-overlapping $k$-month blocks. We define
$$
{\rm elpd}_{\rm WAIC}=-\frac{1}{2}{\rm WAIC},
$$
and report the per-block elpd gain relative to the directly fitted
OU--Binomial baseline at the same aggregation scale:
$$
\Delta {\rm elpd}^{\rm block}_k
=\frac{{\rm elpd}^{\rm model}_k-{\rm elpd}^{\rm OU\text{-}Binomial}_k}{B_k}.
$$

The direct-fitting route should be interpreted as a diagnostic fitting
procedure rather than as a temporally consistent coarse-graining
construction. Because a new latent default-probability path is estimated
independently at each aggregation scale, unresolved within-block latent
default-probability variation can be absorbed either into the fitted latent path
or into the residual covariance parameter. This explains why the covariance
share in Fig.~\ref{fig:direct_fitting_diagnostics}(a) can increase with $k$,
even though the corresponding per-block elpd gain in
Fig.~\ref{fig:direct_fitting_diagnostics}(b) is negative.

\begin{table}[htbp]
\centering
\caption{
Direct-fitting summary for the OU--Vasicek and OU--Davis--Lo extensions.
For each aggregation scale $k$, the models are fitted independently to
non-overlapping $k$-month blocks. The residual parameter is $\rho_{A,k}$ for
OU--Vasicek and $q_k$ for OU--Davis--Lo. The covariance share is the posterior
median fraction of model variance assigned to the residual same-period
covariance component. $\Delta{\rm WAIC}$ and $\Delta{\rm elpd}/B_k$ are
measured relative to the directly fitted OU--Binomial baseline at the same
aggregation scale.
}
\label{tab:direct_fitting_paper_table}
\begin{tabular}{c l c c c c c c}
\hline
$k$ & Model & $q_k$ & $\rho_{A,k}$ & Cov. share
& $\Delta{\rm WAIC}$ & $\Delta{\rm elpd}/B_k$
& Model/data var. \\
\hline
1  & OU--Vasicek   & -- & 0.0025 & 0.041 & 9.01  & -0.009 & 0.963 \\
2  & OU--Vasicek   & -- & 0.0021 & 0.036 & 15.83 & -0.032 & 0.995 \\
3  & OU--Vasicek   & -- & 0.0028 & 0.048 & 25.29 & -0.078 & 0.998 \\
4  & OU--Vasicek   & -- & 0.0026 & 0.042 & 24.32 & -0.100 & 1.000 \\
6  & OU--Vasicek   & -- & 0.0049 & 0.079 & 39.04 & -0.241 & 0.997 \\
12 & OU--Vasicek   & -- & 0.0265 & 0.359 & 69.60 & -0.870 & 1.078 \\
\hline
1  & OU--Davis--Lo & $1.339\times10^{-5}$ & -- & 0.023 & 7.66  & -0.008 & 0.956 \\
2  & OU--Davis--Lo & $2.055\times10^{-5}$ & -- & 0.020 & 11.65 & -0.024 & 0.993 \\
3  & OU--Davis--Lo & $4.046\times10^{-5}$ & -- & 0.027 & 19.33 & -0.059 & 1.003 \\
4  & OU--Davis--Lo & $5.126\times10^{-5}$ & -- & 0.028 & 21.05 & -0.086 & 1.001 \\
6  & OU--Davis--Lo & $1.213\times10^{-4}$ & -- & 0.043 & 30.92 & -0.191 & 0.998 \\
12 & OU--Davis--Lo & $9.383\times10^{-4}$ & -- & 0.145 & 51.59 & -0.645 & 1.000 \\
\hline
\end{tabular}
\end{table}

Some direct fits have relatively low effective sample sizes or mildly elevated
$\hat R$ values, especially for latent OU-state parameters such as the long-run
level $\mu$. This reflects the limited number of non-overlapping blocks and the
weak identifiability of scale-specific latent paths under direct fitting. The
table is therefore used as a diagnostic summary rather than as the primary
basis for structural interpretation.

\clearpage

\section{Renormalized fitting details}
\label{app:renormalized_fit}

This appendix summarizes the renormalized fitting procedure used in
Sec.~III~C. 
The purpose is to separate coarse-grained latent default-probability fluctuations
from residual instantaneous dependence after temporal coarse-graining.

We first fit the OU--Binomial, OU--Davis--Lo, and OU--Vasicek models to the
monthly data at $k=1$. The latent default probability is represented on the
probit scale and follows the same stationary OU/AR(1) specification as in
Appendix~\ref{app:ou_binomial_diagnostics}. The residual-dependence priors are
$$
q\sim {\rm Beta}(2,50), \qquad \rho_A\sim {\rm Beta}(2,8).
$$
From each monthly fit, we extract posterior sample paths of the baseline
monthly default probability,
$$
\{p_t^{(s)}\}_{t=1}^{T},\qquad s=1,\ldots,S .
$$
In the reported implementation, $S=10^3$ posterior paths are used.

For each posterior path, we construct the non-overlapping $k$-month
probabilities using the same survival aggregation as in Sec.~II~B:
$$
p_b^{(k,s)}=1-\prod_{j=0}^{k-1}\{1-p_{bk+j}^{(s)}\}.
$$
The collection
$$
\{p_b^{(k,s)}\}_{b=1,\ldots,B_k;\;s=1,\ldots,S}
$$
defines the model-specific renormalized latent default-probability path at
scale $k$. For $k=1,2,3,4,6,12$, the numbers of non-overlapping blocks are
$$
B_k=489,244,163,122,81,40.
$$

For $k\ge 2$, we do not estimate a new latent OU path. Instead, we condition on
the coarse-grained posterior ensemble and re-estimate only the residual
instantaneous-dependence parameter. Let
$$
\theta_k=q_k \quad \text{for OU--Davis--Lo}, \qquad
\theta_k=\rho_{A,k}\quad \text{for OU--Vasicek}.
$$
For each block $b$, the path-marginalized likelihood is
$$
\widetilde{P}\left(L_b^{(k)}
\mid n_b^{(k)},\theta_k \right)=
\frac{1}{S}\sum_{s=1}^{S}P\left(L_b^{(k)}\mid n_b^{(k)},p_b^{(k,s)},\theta_k \right).
$$
The likelihood for the full $k$-month aggregated data is
$$
\mathcal{L}_k(\theta_k)=\prod_{b=1}^{B_k}
\widetilde{P}\left(L_b^{(k)} \mid n_b^{(k)},\theta_k \right).
$$
The posterior distribution is then
$$
\pi(\theta_k\mid \{L_b^{(k)}\}) \propto \pi(\theta_k)\mathcal{L}_k(\theta_k),
$$
with
$$
q_k\sim {\rm Beta}(2,50),\qquad \rho_{A,k}\sim {\rm Beta}(2,8).
$$
Since only one residual parameter is estimated for each $k$ and model, the
posterior distribution is evaluated on a dense one-dimensional grid, with
higher resolution near zero.

For the renormalized OU--Binomial baseline, no residual parameter is estimated.
The pointwise likelihood is obtained by averaging the binomial likelihood over
the coarse-grained OU--Binomial posterior paths:
$$
\widetilde{P}_{\rm Bin}
\left(L_b^{(k)} \mid n_b^{(k)} \right)=\frac{1}{S}
\sum_{s=1}^{S}{\rm Binomial}\left(L_b^{(k)} \mid n_b^{(k)},p_b^{(k,s)} \right).
$$

\begin{figure}[htbp]
    \centering
    \includegraphics[width=0.90\linewidth]{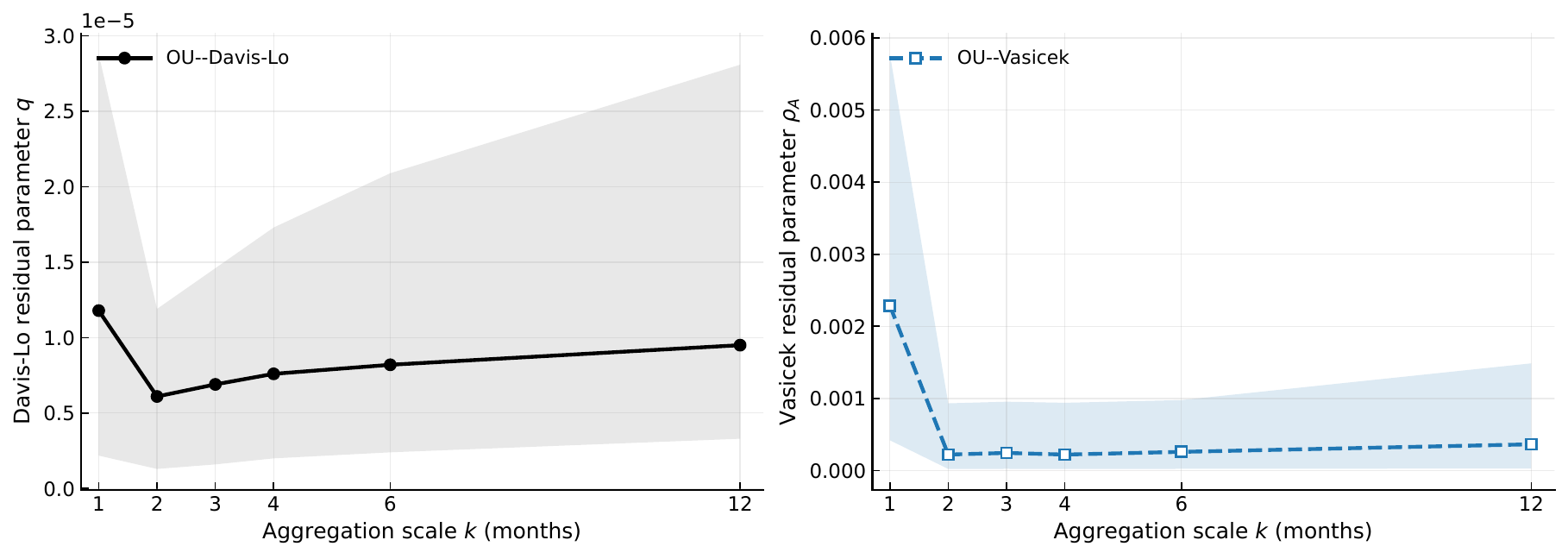}
    \caption{
    Residual-dependence parameters in the renormalized fitting route.
    The left panel shows the Davis--Lo contagion parameter $q_k$, and the
    right panel shows the Vasicek asset-correlation parameter $\rho_{A,k}$.
    For $k=1$, the parameters are obtained from the monthly joint Bayesian
    fits. For $k\ge 2$, they are re-estimated from the path-marginalized
    likelihood conditional on the model-specific coarse-grained posterior
    latent default-probability paths. The Vasicek parameter is sharply reduced
    after temporal coarse-graining, while the Davis--Lo parameter remains
    small.
    }
    \label{fig:app_renormalized_residual_parameters}
\end{figure}

WAIC is computed from the pointwise path-marginalized log-likelihood over
non-overlapping blocks. The corresponding elpd is
$$
{\rm elpd}_{\rm WAIC}=-\frac{1}{2}{\rm WAIC}.
$$
For comparison with Fig.~\ref{fig:renormalized_fitting_diagnostics}(b), we
report the per-block elpd gain relative to the directly fitted OU--Binomial
baseline at the same aggregation scale:
$$
\Delta {\rm elpd}^{\rm block}_k=\frac{{\rm elpd}^{\rm renorm}_{{\rm model},k}
-{\rm elpd}^{\rm direct}_{{\rm OU\text{-}Binomial},k}}{B_k}.
$$
Equivalently,
$$
\Delta{\rm WAIC}_k
={\rm WAIC}^{\rm renorm}_{{\rm model},k}-
{\rm WAIC}^{\rm direct}_{{\rm OU\text{-}Binomial},k}.
$$

The renormalized fitting procedure should be interpreted as a path-based
coarse-graining analysis. It does not ask whether a separately fitted
$k$-month model prefers a residual covariance parameter. 
Instead, it asks whether residual dependence estimated conditional on
temporally coarse-grained monthly posterior paths improves the predictive
description of aggregated default counts.
In the empirical results, the residual covariance share remains small after
coarse-graining, while the per-block elpd gain relative to the directly fitted
OU--Binomial baseline becomes positive for $k\ge 2$; it is also larger than
that of the renormalized OU--Binomial path. 
Thus, the predictive improvement should be interpreted as a consequence of the
scale-consistent conditioning on the coarse-grained latent path, rather than as
evidence for a large residual covariance component. By first accounting for
long-horizon fluctuations through the temporally coarse-grained latent path, the
renormalized route suppresses the over-allocation of variance to residual
instantaneous-dependence parameters. The remaining small residual-dependence
component can then improve the shape of the predictive count distribution.

\begin{table}[htbp]
\centering
\caption{
Renormalized-fitting summary for the OU--Binomial, OU--Vasicek, and
OU--Davis--Lo models. The abbreviations OU--Bin, OU--V, and OU--DL denote
the renormalized OU--Binomial, OU--Vasicek, and OU--Davis--Lo models,
respectively. For OU--V and OU--DL, the residual-dependence parameter is
estimated conditional on the model-specific coarse-grained posterior
latent default-probability path. The residual parameter is $\rho_{A,k}$ for
OU--V and $q_k$ for OU--DL. The intrinsic covariance share is the posterior
median fraction of model variance assigned to the residual same-period
covariance component; it is zero for OU--Bin. $\Delta{\rm WAIC}$ and
$\Delta{\rm elpd}/B_k$ are measured relative to the directly fitted
OU--Binomial baseline at the same aggregation scale. Thus, negative
$\Delta{\rm WAIC}$ and positive $\Delta{\rm elpd}/B_k$ indicate better
WAIC-based predictive fit than the direct OU--Binomial baseline.
The comparison is intended to assess residual dependence conditional on a
scale-consistent coarse-grained latent path, rather than to refit a new latent
process independently at each aggregation scale.
}
\label{tab:renormalized_fitting_paper_table}
\begin{tabular}{c l c c c c c c}
\hline
$k$ & Model & $q_k$ & $\rho_{A,k}$ & Intrinsic cov. share
& $\Delta{\rm WAIC}$ & $\Delta{\rm elpd}/B_k$
& Model/data var. \\
\hline
1  & OU--Bin & -- & -- & 0.000 & 1.56  & -0.002 & 0.946 \\
2  & OU--Bin & -- & -- & 0.000 & 0.21  & -0.000 & 0.996 \\
3  & OU--Bin & -- & -- & 0.000 & 4.98  & -0.015 & 0.993 \\
4  & OU--Bin & -- & -- & 0.000 & -0.43 & 0.002  & 1.008 \\
6  & OU--Bin & -- & -- & 0.000 & -3.52 & 0.022  & 1.006 \\
12 & OU--Bin & -- & -- & 0.000 & 2.77  & -0.035 & 0.990 \\
\hline
1  & OU--V  & -- & $2.196\times10^{-3}$ & 0.040 & 8.76    & -0.009 & 0.958 \\
2  & OU--V  & -- & $2.222\times10^{-4}$ & 0.005 & -122.11 & 0.250  & 0.973 \\
3  & OU--V  & -- & $2.446\times10^{-4}$ & 0.005 & -95.06  & 0.292  & 0.970 \\
4  & OU--V  & -- & $2.147\times10^{-4}$ & 0.004 & -78.85  & 0.323  & 0.995 \\
6  & OU--V  & -- & $2.596\times10^{-4}$ & 0.005 & -59.01  & 0.364  & 0.992 \\
12 & OU--V  & -- & $3.643\times10^{-4}$ & 0.006 & -32.77  & 0.410  & 0.985 \\
\hline
1  & OU--DL & $1.187\times10^{-5}$ & -- & 0.022 & 6.21    & -0.006 & 0.954 \\
2  & OU--DL & $6.100\times10^{-6}$ & -- & 0.007 & -122.26 & 0.251  & 0.945 \\
3  & OU--DL & $6.900\times10^{-6}$ & -- & 0.006 & -95.51  & 0.293  & 0.945 \\
4  & OU--DL & $7.500\times10^{-6}$ & -- & 0.005 & -77.68  & 0.318  & 0.966 \\
6  & OU--DL & $8.100\times10^{-6}$ & -- & 0.004 & -58.32  & 0.360  & 0.972 \\
12 & OU--DL & $9.400\times10^{-6}$ & -- & 0.003 & -32.24  & 0.403  & 0.979 \\
\hline
\end{tabular}
\end{table}

The corresponding numerical summary is reported in Table~\ref{tab:renormalized_fitting_paper_table}.

\clearpage

\begin{acknowledgments}
The author used ChatGPT by OpenAI to assist with manuscript drafting,
English-language editing, and the generation and refinement of data-analysis
code. ChatGPT was not used as an author and was not used to
generate or modify research data or figure images. All analysis code, numerical
results, figures, and scientific interpretations were checked, revised, and
validated by the author, who takes full responsibility for the content of the
manuscript.
\end{acknowledgments}

\section*{Disclosure of interest}

The author declares that there are no competing interests to disclose.

\section*{Funding}

This work was supported by JSPS KAKENHI under Grant JP26K06955.

\section*{Data and Code Availability}

The data analysis and simulation codes used in this study, including all
scripts used to generate the figures and tables, are available in a public
GitHub repository:
\[
\texttt{https://github.com/shintaromori/temporal-renormalization-default-risk}
\]
The empirical default data analyzed in this paper are not included in the
repository, as they are derived from proprietary historical default datasets
and therefore cannot be publicly shared. The repository provides the full
analysis and simulation code so that the results can be reproduced by
researchers with access to comparable data sources. 
It also includes
independently generated low-fidelity synthetic data with the same column
structure and time frequency, provided only for code execution and workflow
demonstration.

\bibliographystyle{apsrev4-2}
\bibliography{references}

@book{MantegnaStanley1999,
  author    = {Mantegna, Rosario N. and Stanley, H. Eugene},
  title     = {An Introduction to Econophysics: Correlations and Complexity in Finance},
  publisher = {Cambridge University Press},
  year      = {1999},
  doi       = {10.1017/CBO9780511755767},
  isbn      = {978-0511755767}
}

@article{Galam2008,
  author  = {Galam, Serge},
  title   = {Sociophysics: A Review of Galam Models},
  journal = {Int. J. Mod. Phys. C},
  volume  = {19},
  number  = {3},
  pages   = {409--440},
  year    = {2008},
  doi     = {10.1142/S0129183108012297}
}

@article{Lux1995,
  author  = {Lux, Thomas},
  title   = {Herd Behaviour, Bubbles and Crashes},
  journal = {Econ. J.},
  volume  = {105},
  number  = {431},
  pages   = {881--896},
  year    = {1995},
  doi     = {10.2307/2235156}
}

@article{LuxMarchesi1999,
  author  = {Lux, Thomas and Marchesi, Michele},
  title   = {Scaling and Criticality in a Stochastic Multi-Agent Model of a Financial Market},
  journal = {Nature},
  volume  = {397},
  pages   = {498--500},
  year    = {1999},
  doi     = {10.1038/17290}
}

@article{Alfarano2005,
  author  = {Alfarano, Simone and Lux, Thomas and Wagner, Friedrich},
  title   = {Estimation of Agent-Based Models: The Case of an Asymmetric Herding Model},
  journal = {Comput. Econ.},
  volume  = {26},
  number  = {1},
  pages   = {19--49},
  year    = {2005},
  month   = {aug},
  doi     = {10.1007/s10614-005-6415-1}
}

@article{Bouchaud2002,
  author  = {Bouchaud, Jean-Philippe and M{\'e}zard, Marc and Potters, Marc},
  title   = {Statistical Properties of Stock Order Books: Empirical Results and Models},
  journal = {Quant. Finance},
  volume  = {2},
  number  = {4},
  pages   = {251--256},
  year    = {2002},
  doi     = {10.1088/1469-7688/2/4/301}
}

@article{FernandezGracia2014,
  author  = {Fernandez-Gracia, J. and Suchecki, K. and Ramasco, J. J. and SanMiguel, M. and Egu{\'i}luz, V. M.},
  title   = {Is the Voter Model a Model for Voters?},
  journal = {Phys. Rev. Lett.},
  volume  = {112},
  pages   = {158701},
  year    = {2014},
  doi     = {10.1103/PhysRevLett.112.158701}
}

@article{MoriHisakadoTakahashi2012,
  author  = {Mori, Shintaro and Hisakado, Masato and Takahashi, Taiki},
  title   = {Phase Transition to a Two-Peak Phase in an Information-Cascade Voting Experiment},
  journal = {Phys. Rev. E},
  volume  = {86},
  pages   = {026109},
  year    = {2012},
  doi     = {10.1103/PhysRevE.86.026109}
}

@article{Mori2019,
  author  = {Mori, Shintaro and Nakayama, Kazuaki and Hisakado, Masato},
  title   = {Voter Model on a Network and Multinomial Distribution},
  journal = {Phys. Rev. E},
  volume  = {99},
  pages   = {052307},
  year    = {2019},
  doi     = {10.1103/PhysRevE.99.052307}
}

@article{SmolyakHavlin2022,
  author  = {Smolyak, Alex and Havlin, Shlomo},
  title   = {Three Decades in Econophysics---From Microscopic Modelling to Macroscopic Complexity and Back},
  journal = {Entropy},
  year    = {2022},
  volume  = {24},
  number  = {2},
  pages   = {271},
  doi     = {10.3390/e24020271}
}

@book{Schonbucher2003,
  author    = {Sch{\"o}nbucher, Philipp J.},
  title     = {Credit Derivatives Pricing Models: Models, Pricing and Implementation},
  publisher = {John Wiley \& Sons},
  year      = {2003}
}

@article{DavisLo2001,
  author  = {Davis, Mark H. A. and Lo, Violet},
  title   = {Infectious defaults},
  journal = {Quant. Finance},
  volume  = {1},
  number  = {4},
  pages   = {382--387},
  year    = {2001},
  doi     = {10.1080/713665832}
}

@article{Vasicek1991,
  author  = {Vasicek, Oldrich A.},
  title   = {Limiting Loan Loss Probability Distribution},
  journal = {KMV Corporation},
  year    = {1991},
  note    = {Working paper}
}

@article{Vasicek2002,
  author  = {Vasicek, Oldrich A.},
  title   = {Loan portfolio value},
  journal = {Risk},
  volume  = {15},
  number  = {12},
  pages   = {160--162},
  year    = {2002}
}

@article{DasDuffieKapadiaSaita2007,
  author  = {Das, Sanjiv R. and Duffie, Darrell and Kapadia, Nikunj and Saita, Leandro},
  title   = {Common Failings: How Corporate Defaults Are Correlated},
  journal = {J. Finance},
  year    = {2007},
  volume  = {62},
  number  = {1},
  pages   = {93--117},
  doi     = {10.1111/j.1540-6261.2007.01202.x}
}

@article{DuffieEcknerHorelSaita2009,
  author  = {Duffie, Darrell and Eckner, Andreas and Horel, Guillaume and Saita, Leandro},
  title   = {Frailty Correlated Default},
  journal = {J. Finance},
  volume  = {64},
  number  = {5},
  pages   = {2089--2123},
  year    = {2009},
  doi     = {10.1111/j.1540-6261.2009.01485.x}
}

@article{AzizpourGieseckeSchwenkler2018,
  author  = {Azizpour, S. and Giesecke, K. and Schwenkler, G.},
  title   = {Exploring the Sources of Default Clustering},
  journal = {J. Financ. Econ.},
  volume  = {129},
  number  = {1},
  pages   = {154--183},
  year    = {2018},
  doi     = {10.1016/j.jfineco.2018.04.008}
}

@article{SakataHisakadoMori2007,
  author  = {Sakata, Ayaka and Hisakado, Masato and Mori, Shintaro},
  title   = {Infectious Default Model and Recovery Rate},
  journal = {J. Phys. Soc. Jpn.},
  volume  = {76},
  number  = {5},
  pages   = {054801},
  year    = {2007},
  doi     = {10.1143/JPSJ.76.054801}
}

@article{TorriGiacomettiFarina2026,
title = {Modeling portfolio loss distribution under infectious defaults and immunization},
journal = {Commun. Nonlinear Sci. Numer. Simul.},
volume = {159},
pages = {109886},
year = {2026},
issn = {1007-5704},
doi = {https://doi.org/10.1016/j.cnsns.2026.109886},
url = {https://www.sciencedirect.com/science/article/pii/S1007570426002467},
author = {Gabriele Torri and Rosella Giacometti and Gianluca Farina}
}

@article{HisakadoMori2021,
  title   = {Parameter estimation of default portfolios using the Merton model and phase transition},
  journal = {Physica A: Statistical Mechanics and its Applications},
  volume  = {563},
  pages   = {125435},
  year    = {2021},
  issn    = {0378-4371},
  doi     = {10.1016/j.physa.2020.125435},
  url     = {https://www.sciencedirect.com/science/article/pii/S0378437120307615},
  author  = {Hisakado, Masato and Mori, Shintaro}
}

@article{Hawkes1971,
  author  = {Hawkes, A. G.},
  title   = {Spectra of Some Self-Exciting and Mutually Exciting Point Processes},
  journal = {Biometrika},
  volume  = {58},
  number  = {1},
  pages   = {83--90},
  year    = {1971},
  doi     = {10.1093/biomet/58.1.83}
}

@article{Kirchner2017,
  author  = {Kirchner, Matthias},
  title   = {An estimation procedure for the Hawkes process},
  journal = {Quant. Finance},
  volume  = {17},
  number  = {4},
  pages   = {571--595},
  year    = {2017},
  doi     = {10.1080/14697688.2016.1211312}
}

@article{ErraisGieseckeGoldberg2010,
  author  = {Errais, Eymen and Giesecke, Kay and Goldberg, Lisa R.},
  title   = {Affine Point Processes and Portfolio Credit Risk},
  journal = {SIAM J. Financial Math.},
  volume  = {1},
  number  = {1},
  pages   = {642--665},
  year    = {2010},
  doi     = {10.1137/090771272}
}

@article{HisakadoHattoriMori2022,
  title = {From the multiterm urn model to the self-exciting negative binomial distribution and Hawkes processes},
  author = {Hisakado, Masato and Hattori, Kodai and Mori, Shintaro},
  journal = {Phys. Rev. E},
  volume = {106},
  issue = {3},
  pages = {034106},
  numpages = {9},
  year = {2022},
  month = {Sep},
  publisher = {American Physical Society},
  doi = {10.1103/PhysRevE.106.034106},
  url = {https://link.aps.org/doi/10.1103/PhysRevE.106.034106}
}

@misc{Mori2026ContagionMacro,
  author        = {Mori, Shintaro},
  title         = {Contagion or Macroeconomic Fluctuations? Identifiability in Aggregated Default Data},
  year          = {2026},
  eprint        = {2604.18118},
  archivePrefix = {arXiv},
  primaryClass  = {q-fin.RM},
  doi           = {10.48550/arXiv.2604.18118},
  note          = {arXiv preprint arXiv:2604.18118}
}

\end{document}